\documentclass[aps,twocolumn,prb,showpacs,superscriptaddress]{revtex4}
\usepackage{amssymb}
\usepackage{epsfig}
\usepackage{epsf}
\usepackage{subfigure}

\def\frac#1#2{{\textstyle{#1 \over #2}}}

\def\be{\begin{equation}} \def\ee{\end{equation}}
\def\bea{\begin{eqnarray}} \def\eea{\end{eqnarray}}

\def\nn{\nonumber}

\def\ssr{\scriptscriptstyle\rm}

\def\yd{^\dagger}
\def\nd{^{\vphantom{\dagger}}}

\def\kvec{{\vec k}}

\def\Qvec{{\vec Q}}

\begin{document}

\title{Microscopic Theory of the Thermodynamic Properties of Sr$_3$Ru$_2$O$_7$}

\author{Wei-Cheng Lee}
\email{leewc@physics.ucsd.edu}

\author{Congjun Wu}
\email{wucj@physics.ucsd.edu}
\affiliation{Department of Physics, University of California, San Diego, California, 92093, USA}

\date{\today}

\begin{abstract}
The thermodynamic properties of the bilayer Sr$_3$Ru$_2$O$_7$ at very low temperatures are investigated by 
a realistic tight-binding model with the on-site interactions treated at the mean-field level.
Due to the strong spin-orbit coupling, the band structure undergoes 
a significant change in Fermi surface topology as the external magnetic 
field is applied, invalidating the rigid band picture in which the Zeeman 
energy only causes chemical potential shifts. 
In addition, since Sr$_3$Ru$_2$O$_7$ is a $t_{2g}$ active system with
unquenched orbital moments, the orbital Zeeman energy is not 
negligible and plays an important role in the phase diagram on the 
magnetic field orientation.
We find that both the total density of states at the Fermi energy and 
the entropy exhibit a sudden increase near the critical magnetic field 
for the nematic phase, echoing the experimental findings.
Our results suggest that extra cares are necessary to isolate the 
contributions due to the quantum criticality from the 
band structure singularity in this particular material. 
The effects of quantum critical fluctuations are briefly discussed.
\end{abstract}

\maketitle

\section {Introduction}

Recently, a great deal of attention has been paid to the bilayer 
ruthenate compound Sr$_3$Ru$_2$O$_7$ with various interesting physical 
properties.
It was  first considered as a field-tuned quantum critical state having
metamagnetic transitions around 8 Tesla\cite{grigera2001,perry2001,grigera2003}.
Later, in the ultra-pure single crystal it has been found that the 
metamagnetic quantum critical point is intervened by 
the emergence of an unconventional anisotropic (nematic) electronic 
state\cite{grigera2004,borzi2007}, stimulating considerable theoretical
efforts \cite{millis2002, green2004,kee2005,yamase2007,yamase2007a,
puetter2007,berridge2009,berridge2010, lee2009nematic,raghu2009,puetter2010,
fischer2010,lee2010nematic}.
Sr$_3$Ru$_2$O$_7$ is a metallic itinerant system with the active 
$t_{2g}$-orbitals of the Ru sites in the bilayer RuO$_2$ $(ab)$ planes.
At very low temperatures ($\sim$ 1K),
it starts as a paramagnet at small magnetic fields.
Further increasing field strength leads to two consecutive 
metamagnetic transitions at 7.8 and 8.1 
Tesla if the field is perpendicular to the $ab$-plane.
The nematic phase is observed between these two transitions, 
identified by the observation of anisotropic resistivity 
without noticeable lattice distortions.

This nematic phase in Sr$_3$Ru$_2$O$_7$ can be understood as the particle-hole 
channel Fermi surface instability of the Pomeranchuk-type
\cite{grigera2003}.
It is a mixture of the $d$-wave Pomeranchuk instabilities in both 
density and spin channels \cite {wu2007}, 
though the microscopic origin of these instabilities remain controversial.
Different microscopic theories have been proposed based 
on the quasi-1D bands of $d_{xz}$ and $d_{yz}$\cite{lee2009nematic,
raghu2009,lee2010nematic}, and based on the 2d-band of $d_{xy}$
\cite{kee2005,yamase2007,puetter2010,fischer2010}.
In our\cite{lee2009nematic,lee2010nematic} and Raghu {\it et. al.}'s 
\cite{raghu2009} theories, the unconventional (nematic) magnetic 
ordering was interpreted as orbital ordering among the $d_{xz}$ 
and $d_{yz}$-orbitals.
In particular, in Ref. [\onlinecite{lee2010nematic}] a realistic 
tight-binding model with the important 
features including the $t_{2g}$-orbitals, the bilayer splitting, spin-orbit 
coupling, and the staggered rotations of RuO octahedra,
has been derived and shown to reproduce accurately the results of 
the angle-resolved  photon emission spectroscopy (ARPES) 
\cite{tamai2008} and the quasiparticle interference in the 
spectroscopic imaging scanning tunneling microscopy (STM) \cite{lee2009qpi327}.

On the other hand, orbital physics is an important subject in  
transition metal oxides with active $d$-orbitals\cite{salamon2001,
khaliullin2005}. 
Recently the research of orbital physics has been further stimulated due 
to the observations of anisotropic properties in iron-pnictides
\cite{zhao2009,chuang2010} which are consistent with orbital 
ordering \cite{lee2009qpi,lv2010,chen2010}. 
Generally speaking, a transition metal oxide with active $t_{2g}$-orbitals 
could exhibit orbital ordering in the quasi-1D bands under 
certain conditions.
In the particular case of Sr$_3$Ru$_2$O$_7$, it has been argued 
\cite{lee2009nematic,raghu2009} that at zero magnetic field, although 
the system does not show any order, 
it is near the instabilities of ferromagnetic and orbital 
(nematic) orderings.
Applying the magnetic field pushes the system closer to the van Hove
singularities which largely raises 
up the density of states at the Fermi energy ${\cal D}(\epsilon_F)$ 
and results in both the metamagnetic and nematic transitions. 
It should be noted that the Fermi liquid states always dominate the 
low temperature physics of this system because the resistivity was
found to follow $T^2$ law in both isotropic and nematic phases. 
This suggests that quantum criticality is not expected to play much 
role in both isotropic and nematic phases at low temperatures.
Nevertheless, quantum critical behavior appears at finite 
temperatures above the nematic phase, which exhibits
non-Fermi liquid properties.


The study of quantum critical states has attracted intensive research 
activities over past few decades\cite{coleman2005,lohneysen2007}.  
Materials possessing quantum critical points are usually triggered 
by tunable parameters, for example, external pressure and magnetic field. 
When approaching quantum critical points, 
not only the zero temperature electronic ground state but also the 
low-temperature properties are modified radically.
Pioneering theoretical works done by Hertz\cite{hertz1976} and extended 
by Millis\cite{millis1993} have addressed the physics
near the ferromagnetic and antiferromagnetic quantum critical points, 
showing that the critical fluctuations
are scale invariant in the vicinity of  quantum critical points due
to the divergence of the correlation length right at these points.
This aspect enables the establishment of universal classes based on 
the scaling theory.
Experimentally, quantum critical states usually exhibit divergences 
in thermodynamic properties like specific heat, entropy, {\it etc}.
The critical exponents associated with these divergences are 
believed to be universal values in the same universal class regardless 
the microscopic details.

The influence of quantum critical fluctuations in Sr$_3$Ru$_2$O$_7$
seems to be novel as well.
Rost {\it et. al.}\cite{rost2009,rost2010} measured the
entropy and specific heat in ultra-pure samples and 
found divergences near the metamagnetic transitions in both quantities.
Although it is a common feature in a quantum critical state that the 
specific heat $C$ diverges as $C/T=A[(H-H_c)/H_c]^{-\alpha}+B$ due to 
quantum fluctuations, the exponent of $\alpha$ is fitted 
to be 1 instead of $1/3$ as predicted by the Hertz-Millis theory
\cite{hertz1976,millis1993}.
The total density of states (DOS) measured by Iwaya {\it et al.}
\cite{iwaya2007} using the STM showed that the DOS at the Fermi 
energy (${\cal D}(\epsilon_F)$)
increases significantly under the application of the magnetic field,
but the DOS at higher and lower energy does not change accordingly.
This indicates that the Zeeman energy in this case does not simply 
cause a chemical potential shift so that the conventional rigid 
band picture can not explain this result.
These findings have posted a challenge to understand the critical 
behavior in this material.

In this article, we show that the tight-binding model derived by 
Arovas and two of us\cite{lee2010nematic} together 
with the on-site interactions treated simply at the mean-field 
level already gives rise to divergences in these properties without 
involving quantum fluctuations.
Realistic features like multi-orbital bands, the bilayer splitting, 
the rotations of RuO octahedra, and spin-orbit coupling
make this material very sensitive to small energy scales. 
Because parts of the Fermi surface are close to the van Hove singularities, 
Fermi surface reconstructions in the external magnetic fields 
lead to a singular behavior in ${\cal D}(\epsilon_F)$.
This results in the divergences observed in the experiments mentioned above.
Because of the strong spin-orbit coupling and the unquenched
orbital moments, the Zeeman energy tends to 
reconstruct the Fermi surfaces rather than just provides a chemical 
potential shift.
Our results suggest that the influence of  quantum critical 
fluctuations will be masked if ${\cal D}(\epsilon_F)$ of the system exhibits a 
non-monotonic behavior with the tuning parameters, implying that a 
more careful analysis is required in order to distinguish the role of the 
quantum criticality in the bilayer Sr$_3$Ru$_2$O$_7$.

This paper is organized as follows.
We will summarize the tight-binding model derived in Ref. 
[\onlinecite{lee2010nematic}] in Sect. 
\ref{sect:TB}.
The formalism of the mean-field theory will be 
presented in Sect. \ref{sect:MFT}.
The results for the cases of the magnetic fields parallel to and tilted
away from the $c$-axis will be discussed in Sect. \ref{sect:Bperp} and 
\ref{sect:Btilt}, respectively.
Conclusions will be given in Sect. \ref{sect:cln}.


\section{The tight-binding model}
\label{sect:TB}
The detailed band structure of Sr$_3$Ru$_2$O$_7$ is complicated by
the $t_{2g}$-orbital structure (e.g. $d_{xz}, d_{yz}, d_{xy}$), 
the bilayer splitting, the staggered distortion of the RuO octahedra, 
and spin-orbit coupling.
We have constructed a detailed tight-binding Hamiltonian which
gives rise to band structures in agreement with the ARPES data
in a previous work \cite{lee2010nematic}.
We found that a difference of the on-site potential between the two 
adjacent RuO layers, $V_{\rm bias}$, should be added \cite{lee2010nematic}
in order to fit the shape of the Fermi surfaces observed in the ARPES 
experiments \cite{tamai2008}.
This term appears because ARPES is a surface probe and this bilayer
symmetry breaking effect is important near the surface.
Since we focus on the thermodynamic properties which are all bulk properties, 
$V_{\rm bias}$ is set to be zero in this paper.
Below we will start from this model and refer readers to
Ref. \onlinecite{lee2010nematic} for more detailed information.

The tight-binding band Hamiltonian $H_0$ can be reduced 
to block forms classified by $k_z=0,\pi$ corresponding to bonding and 
anti-bonding bands with respect to layers as:
\bea
H_0 &=& h_0(k_z=0) + h_0(k_z=\pi),
\label{eq:h0}
\eea
with $h_0(k_z)$ defined as
\be
h_0(k_z) = {\sum_\kvec}' \Phi^{\dagger}_{\kvec,k_z,s}
\left(
\begin{array}{cc}
\hat{h}_{0s}(\kvec,k_z)& \hat{g}^\dagger(\kvec,k_z)\\
\hat{g}(\kvec,k_z)&\hat{h}_{0s}(\kvec+\Qvec,k_z)\\
\end{array}\right) \Phi_{\kvec,k_z,s},  \nn \\
\label{eq:h0kz}
\ee
where the spinor $\Phi^{\dagger}_{\kvec,k_z,s}$ operator is defined as
\bea
\Phi^{\dagger}_{\kvec,k_z,s}
&=&\big( d^{yz\, \dagger}_{\kvec,s,k_z} \,,\,
d^{xz\,\dagger}_{\kvec,s,k_z} \,,\,
d^{xy\,\dagger}_{\kvec,-s,k_z} \,,\,  \nn \\
&&  d^{yz\,\dagger}_{\kvec+\Qvec,s,k_z} \,,\, d^{xz\,\dagger}_{\kvec+\Qvec,s,k_z} \,,\,
d^{xy\,\dagger}_{\kvec+\Qvec,-s,k_z} \big);
\label{eq:spinor}
\eea
$d^{\alpha}_{s,k_z} (\vec k)$ annihilates an electron with 
orbital $\alpha$ and spin polarization $s$ at momentum $(\kvec,k_z)$;
$\Qvec=(\pi,\pi)$ is the nesting wavevector corresponding to unit 
cell doubling induced by the rotations of RuO octahedra;
${\sum_{\kvec}}'$ means that only half of the Brillouin zone is summed.
Please note the opposite spin configurations $s$ and -$s$ for 
the $d_{xz}, d_{yz}$ and $d_{xy}$-orbitals in Eq. \ref{eq:spinor},
which is convenient for adding spin-orbit coupling later.

The diagonal matrix kernels $\hat{h}_{0s}$ in Eq. \ref{eq:h0kz} are defined 
as
\bea
\hat{h}_{0s}(\kvec,k_z) &=& \hat{A}_s(\kvec) + \hat{B}_1 \cos k_z 
- \mu\hat{I},
\eea
where
\bea
\hat{A}_s(\vec k)&=&
\left(\begin{array}{ccc}
\epsilon_{\kvec}^{yz}& \epsilon_{\kvec}^{\rm off}+is\lambda & ~~-s\lambda\\
\epsilon_{\kvec}^{\rm off}-is\lambda & \epsilon_{\kvec}^{xz} & ~~i\lambda\\
-s\lambda & -i \lambda &~~ \epsilon_{\kvec}^{xy}
\end{array}\right),
\label{eq:as}
\eea
and
\bea
\hat{B}_1&=&\left(\begin{array}{ccc}
-t_\perp&0&0\\ 0&-t_\perp&0\\ 0&0&0
\end{array} \right);
\eea
where $t_\perp$ is the longitudinal inter-layer hopping for the $d_{xz}$ and $d_{yz}$ orbitals.
$\lambda$ is the spin-orbit coupling strength
which comes from the on-site spin-orbit coupling term as
$H_{so}=\lambda\sum_i \vec L_i \cdot \vec S_i$; 
$\mu$ is the chemical potential;
the dispersions for the $d_{yz}$, $d_{xz}$, and $d_{xy}$ bands 
in Eq. \ref{eq:as} are defined as
\bea
\epsilon^{yz}_{\kvec} &=& - 2 t_2 \cos k_x - 2 t_1 \cos k_y, \nn \\
\epsilon^{xz}_{\kvec} &=&
  - 2 t_1 \cos k_x - 2 t_2 \cos k_y, \nn \\
  \epsilon^{xy}_{\kvec} &=& - 2 t_3 \big( \cos k_x 
  + \cos k_y \big) - 4 t_4 \cos k_x \cos k_y \nn \\
 && -2 t_5 \big( \cos 2 k_x + \cos 2 k_y \big) - V_{xy} \nn \\
\epsilon^{\rm off}_{\kvec} &=& - 4 t_6 \sin k_x \sin k_y,
\eea
which includes longitudinal ($t_1$) and transverse ($t_2$) hopping for the
the $d_{xz}$ and $d_{yz}$ orbitals, respectively, as well as
are nearest neighbor ($t_3$), next-nearest neighbor ($t_4$), and next-next-nearest
neighbor ($t_5$) hopping for the $d_{xy}$ orbital. 
Following the previous LDA calculations \cite{singh2001},
$V_{xy}$ is introduced to account for the splitting of the $d_{yz}$ and $d_{xz}$ states relative to the $d_{xy}$ states.
While symmetry forbids nearest-neighbor hopping between different $t_{2g}$ orbitals 
in a perfect square lattice without the rotation of Ru octahedra, a term describing hopping
between $d_{xz}$ and $d_{yz}$ orbitals on next-nearest neighbor sites ($t_6$) is allowed and put into the tight-binding model.

The off-diagonal matrix kernel $\hat{g}(\kvec,k_z)$ in Eq. \ref{eq:h0kz}
reads
\bea
\hat{g}(\kvec,k_z) &=& \hat{G}(\kvec) - 2\hat{B}_2 \cos k_z,
\eea
where

\bea
\hat{B}_2 &=&\left(
\begin{array}{ccc}
0&t^\perp_{\ssr{INT}}&0\\
-t^\perp_{\ssr{INT}}&0&0\\
0&0&0
\end{array}\right),
\eea
and
\bea
\hat{G}(\kvec)&=& \left(
\begin{array}{ccc}
0&-2t_{\ssr{INT}}\,\gamma(\vec k)&0\\
2t_{\ssr{INT}}\,\gamma(\vec k) &0&0\\
0&0&0\end{array}\right),
\eea
with $\gamma(\vec k)=\cos k_x +\cos k_y$.
$t_{\ssr{INT}}$ and $t^\perp_{\ssr{INT}}$ describe the intra- and inter-layer hopping between $d_{xz}$ and $d_{yz}$ 
induced by the rotations of RuO octahedra, providing the coupling between $\vec{k}$ and $\vec{k}+\vec{Q}$.

When describing the Zeeman energy, we can choose the magnetic field 
$\vec{B}$ to lie on the $xz$ plane and define $\theta$ 
as the angle between $\vec{B}$ and the $c$-axis of the sample without loss of the generality.
Consequently, the Zeeman term becomes:
\bea
H_{Zeeman} &=& H^{orbital}_{Zeeman} + H^{spin}_{Zeeman},\nn\\
H^{orbital}_{Zeeman}&=&-\mu_B\,B\sum_{i,a} \big(L_{z,ia}
\cos\theta+L_{x,ia}\sin\theta\big),\nn\\
H^{spin}_{Zeeman}&=&-2\mu_B\,B\sum_{i,a,\alpha} 
\big(S^\alpha_{z\,ia}\cos\theta+S^\alpha_{x\,ia}\sin\theta\big), \nn \\
\eea
where $a$ is the layer index, $B=\vert\vec{B}\vert$, and the matrices $L_{x,z}$ 
can be found in Ref. [\onlinecite{lee2010nematic}].

For $\theta\neq 0$, the extra Zeeman terms from $x$-component 
of $\vec{L}$ and $\vec{S}$ couple $\Phi^{\dagger}_{\kvec,k_z,\uparrow}$ 
and $\Phi^{\dagger}_{\kvec,k_z,\downarrow}$.
Defining $\phi\yd_{\kvec,k_z}\equiv \big(\Phi^\dagger_{\kvec,k_z,\uparrow},
\Phi^\dagger_{\kvec,k_z,\downarrow}\big)$, the Zeeman term can be written 
in the matrix form as:
\bea
H_{Zeeman}&=&\mu_B B{\sum_{\kvec}}'\sum_{k_z}\phi\yd_{\kvec,k_z}
\hat{H}_Z(\theta)\phi\nd_{\kvec,k_z},\nn\\
\eea
where 
\be
\hat{H}_Z(\theta)= \left(
\begin{array}{cccc}
\hat{H}^D_Z(\theta,+)&0&\hat{H}^{O\,\dagger}_Z(\theta)&0\\
0&\hat{H}^D_Z(\theta,+)&0&\hat{H}^{O\,\dagger}_Z(\theta)\\
\hat{H}^O_Z(\theta)&0&\hat{H}^D_Z(\theta,-)&0\\
0&\hat{H}^O_Z(\theta)&0&\hat{H}^D_Z(\theta,-)
\end{array}\right),
\ee
\be
\hat{H}^D_Z(\theta,s)= \cos\theta\times\left(
\begin{array}{ccc}
-s&-i&0\\
i&-s&0\\
0&0&s
\end{array}\right),
\ee
with $s=\pm 1$ and
\be
\hat{H}^O_Z(\theta)= \sin\theta\times\left(
\begin{array}{ccc}
-1&0&0\\
0&-1&-i\\
0&i&-1
\end{array}\right),
\ee

\section{The Mean-field Theory}
\label{sect:MFT}

\begin{figure}
\includegraphics[width=0.8\linewidth]{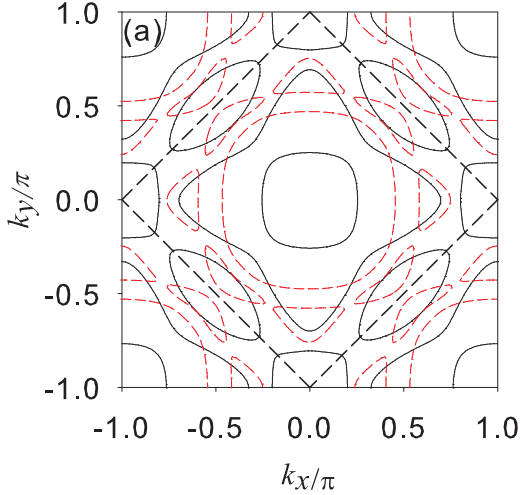}
\caption{\label{fig:fsb0} (Color online) The Fermi surfaces using the 
bilayer tight-binding model Eq. \ref{eq:h0kz} with the 
parameters in unit of $t_1$ as: $t_{2} = 0.1t_1, t_{3} = t_1, t_{4} = 0.2t_1,
t_{5} = -0.06t_1, t_{6} = 0.1t_1,t_\perp=0.6t_1, t_{\ssr{INT}}=t^\perp_{\ssr{INT}}=0.1t_1, \lambda = 0.2t_1,
V_{xy}=0.3t_1$, and $\mu=0.94t_1$.
The thick dashed lines mark the boundary of half Brillouin zone due
to the unit cell doubling induced by the rotation of RuO octahedra. 
The Fermi surfaces of the bonding ($k_z=0$, black 
solid lines) and the anti-bonding bands ($k_z=\pi$, red dashed lines) 
could cross since $k_z$ is a good quantum number.}
\end{figure}

In Sect. \ref{sect:TB}, we have introduced the rather complicated
band structure of Sr$_3$Ru$_2$O$_7$.
In this section, we introduce the minimal multi-orbital band Hubbard model
for the nematic ordering and outline the mean-field procedure.

In realistic band structures measured by ARPES \cite{tamai2008},
there exists an additional $\delta$-band arising from the 
$d_{x^2-y^2}$-orbital which is not covered by the current model.
The particle filling in the $t_{2g}$-orbitals is not fixed.
For the convenience of calculation, we fix the chemical potential $\mu=0.94t_1$ 
instead of fixing particle filling in the $t_{2g}$-orbitals while
changing magnetic fields and orientations in Sect. \ref{sect:Bperp}
and Sect. \ref{sect:Btilt}.
The corresponding fillings inside the $t_{2g}$-orbitals per Ru atom varies
within the range between 4.05 and 4.06 in Figs. \ref{fig:phase-0}, \ref{fig:phase-T}, \ref{fig:phase}.
This treatment does not change any essential qualitative physics.

The Hubbard model contains the on-site intra and 
inter orbital interactions as
\bea
H_{int} = U\sum_{i,a,\alpha} \hat{n}^\alpha_{ia\uparrow}\hat{n}^\alpha_{ia\downarrow} 
+ \frac{V}{2}\sum_{i,a,\alpha\neq\beta} \hat{n}^\alpha_{ia}\,\hat{n}^\beta_{ia},
\eea
where the Greek index $\alpha$ refers to the orbitals $xz,yz$ and $xy$;
the Latin index $a$ refers to the upper and lower layers.
Throughout this paper, the parameter values are taken as
$U/t_1=V/t_1=3.6$.
The other two possible terms in the multi-band Hubbard interaction 
are the Hund's rule coupling and pairing hopping terms, which
do not change the qualitative physics and are neglected.
We assume the external $B$-field lying in the $xz$-plane with
an angle $\theta$ tilted from the $z$-axis.
The occupation and spin in each orbital and layer are
defined as follows:
\bea
n^\alpha_a&\equiv& \sum_s \langle d^{\alpha\,\dagger}_{s,a}(i)
d^{\alpha}_{s,a}(i)\rangle,\nn\\
S^\alpha_{z\,a}&\equiv& \frac{1}{2}\sum_s s\,\langle 
d^{\alpha\,\dagger}_{s,a}(i)d^{\alpha}_{s,a}(i)\rangle,\nn\\
S^\alpha_{x\,a}&\equiv& \frac{1}{2}\sum_s 
\langle d^{\alpha\,\dagger}_{s,a}(i)d^{\alpha}_{\bar s,a}(i)\rangle.
\eea
No any other inter-layer interaction is considered and the non-interacting
Hamiltonian has symmetric layers, thus the order parameters are the 
same for both layers, which have been numerically confirmed.
From now on it will be assumed: 
\bea
n^\alpha_a &=& n^\alpha, \ \ \ 
S^\alpha_{x,z\,a}=S^\alpha_{x,z}.
\eea
With this property, $k_z$ remains a good quantum number in 
the resulting mean-field Hamiltonian.

The standard mean-field decomposition of $H_{int}$ leads to 
\be
H^{MF}_{int} = \sum_{i,a,\alpha}\sum_s  W^\alpha_{s} 
d^{\alpha\,\dagger}_{s,a}(i)d^{\alpha}_{s,a}(i)
-U S^\alpha_{x} d^{\alpha\,\dagger}_{s,a}(i)d^{\alpha}_{\bar{s},a}(i)\,,
\label{eq:mfint}
\ee
where
\be
W^\alpha_{s} = U\big( \frac{1}{2}n^\alpha-s\,S^\alpha_{z}\big) 
+ V\sum_{\beta\neq\alpha}n^\beta.
\ee
The interaction parameters $(U,V)$ are chosen such that no spontaneous
magnetization occurs in the absence of the external magnetic field.
Moreover, since the order parameters $\{n^\alpha_a\}$ are non-zero 
even without magnetic field, we require that the renormalized Fermi surface 
at zero field to be the one given in Fig. \ref{fig:fsb0}.
As a result, in addition to the optimized parameters obtained in 
our previous work\cite{lee2010nematic}, we need to subtract the
following term from Eq. \ref{eq:mfint}:
\bea
H_{shift} &=& \sum_{i,a,\alpha}\sum_s  W^\alpha_{s}(0) d^{\alpha\,\dagger}_{s,a}(i)d^{\alpha}_{s,a}(i),
\eea
where 
\bea
W^\alpha_{s}(0) &=& \frac{1}{2}U n^\alpha(0) + V\sum_{\beta\neq\alpha}n^\beta(0),
\eea
and $n^\alpha(0)$ is the occupation number in orbital $\alpha$ 
corresponding to the Fermi surfaces shown in Fig. \ref{fig:fsb0}.
This is an effect of the renormalization of the chemical potential $\mu$
and $V_{xy}$ due to interactions.
After putting all the pieces together, we finally arrive at 
the mean-field Hamiltonian as:
\bea
H^{MF}&=&H_0 + H^{MF}_{int}-H_{shift} + H_{Zeeman}\nn\\
&\equiv& {\sum_{\kvec}}'\sum_{k_z} \phi\yd_{\kvec,k_z} \, H^{MF}(\vec{k}) \phi\nd_{\kvec,k_z}.
\label{mfall}
\eea

The order parameters are computed self-consistently. 
It has been pointed out\cite{lee2009nematic,raghu2009} that 
the nematic phase can be identified as the orbital ordering 
between the $d_{xz}$ and $d_{yz}$-orbitals, thus
the nematic order parameter ${\cal N}$ is defined as
\bea
{\cal N} = 2(n^{yz} - n^{xz}).
\label{eq:N}
\eea
The magnetization order parameter is defined as:
\bea
\vec{{\cal M}}\equiv 2\sum_{\alpha} \vec{S}^\alpha.
\label{eq:M}
\eea
The factor of 2 in Eq. \ref{eq:M} and \ref{eq:N} accounts
for the double layers.

\section{The Case of the perpendicular magnetic field
($\theta=0$)}
\label{sect:Bperp}

In this section the thermodynamic properties at low temperatures
are calculated within the mean-field theory for the case 
of $\vec{B}\parallel \hat{c}$, {\it i.e.,} $\theta=0$.
Most important features of these properties, including the 
evolution of the nematic ordering, the tunneling DOS, 
entropy landscapes, and the finite temperature phase diagram 
under the application of the magnetic field $B$,
can be reasonably reproduced and understood 
by the singular behavior of ${\cal D}(\epsilon_F)$ under the magnetic field.

\subsection{Nematic ordering and metamagnetic transitions}

\begin{figure}
\includegraphics[width=0.8\linewidth]{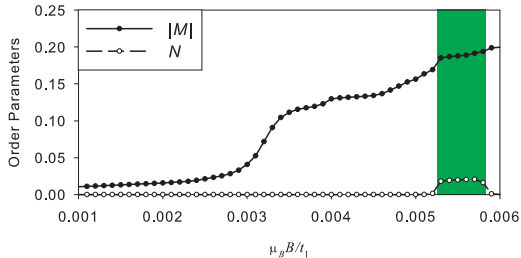}
\caption{\label{fig:phase-0} (Color online) The order parameters as 
a function of $\mu_B B$ for $\theta=0$.
The nematic phase (in the green area) is bounded by two magnetization 
jumps which corresponds to metamagnetic transitions.}
\end{figure}

\begin{figure*}
	\begin{center}
	\subfigure{
	\epsfig{figure=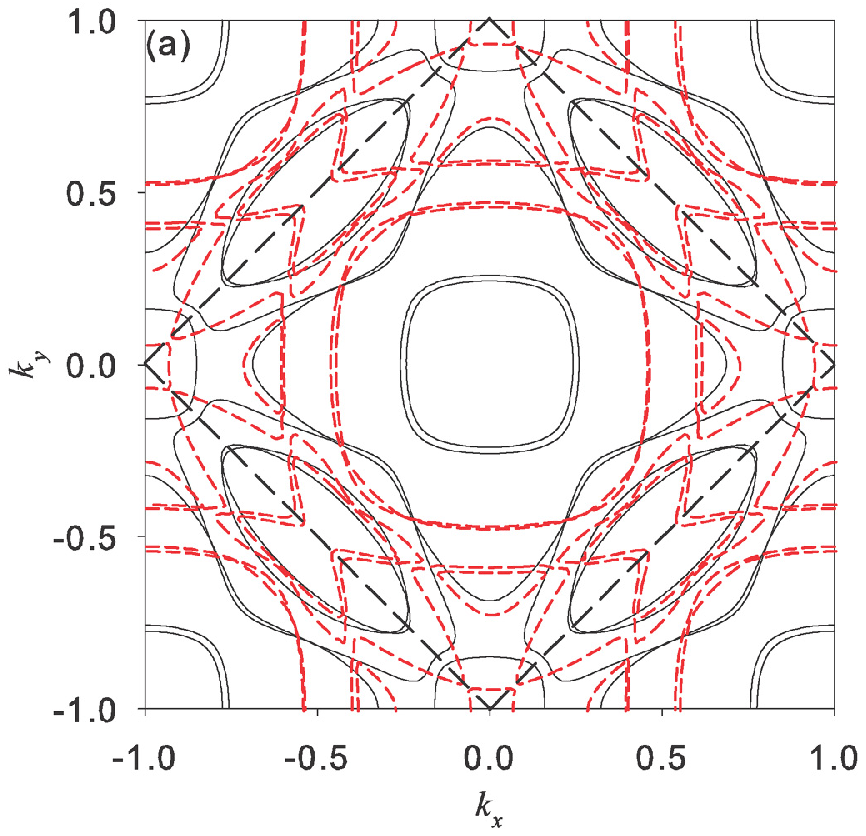,width=0.3\linewidth}}
        \subfigure{
        \epsfig{figure=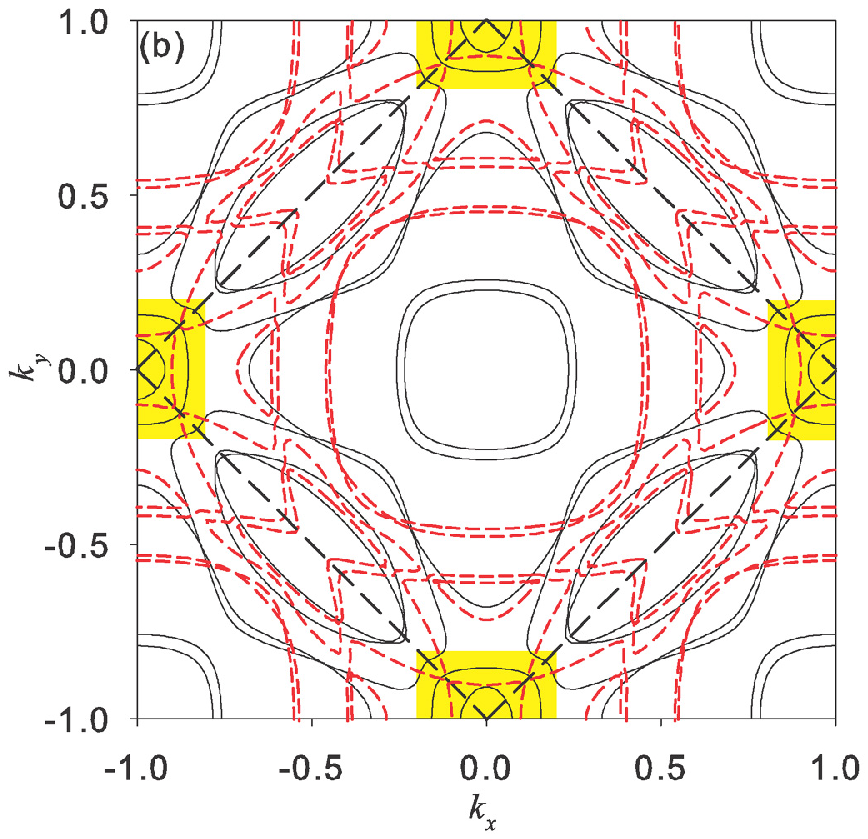,width=0.3\linewidth}}
	\subfigure{
	\epsfig{figure=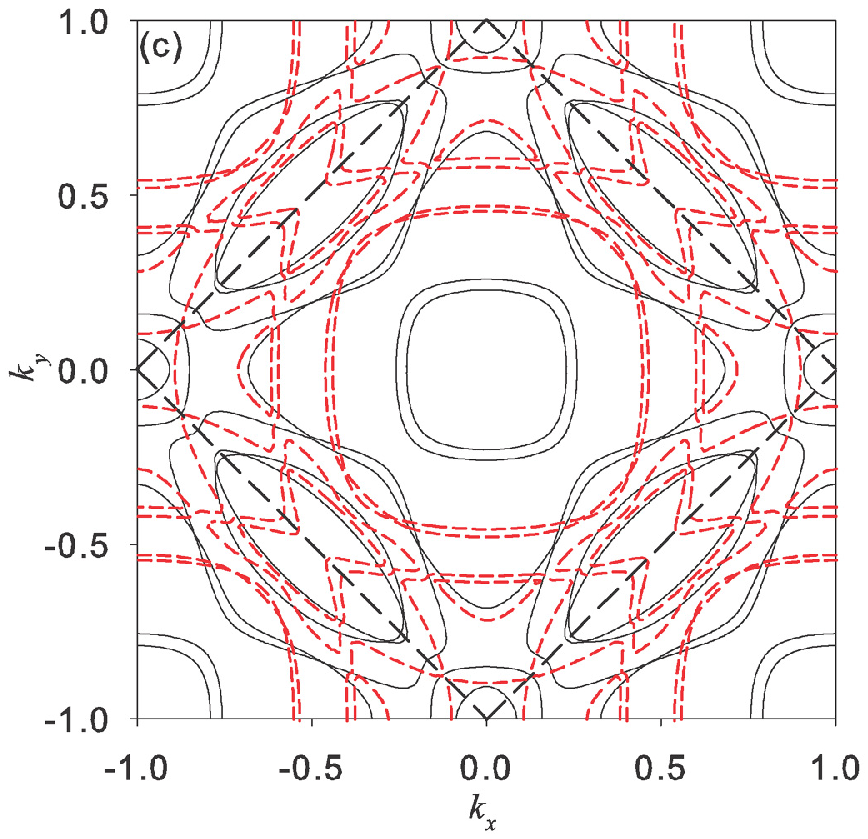,width=0.3\linewidth}}
	\end{center}
	\caption{Fermi surface evolution for the case of magnetic 
field parallel to $c$ axis 
        (a) before ($\mu_B B=0.0048 t_1$) (b) inside ($\mu_B B=0.00544 t_1$), and (c)after 
($\mu_B B=0.006 t_1$) the nematic phase. 
        Significant changes in the Fermi surface topology under
 the magnetic field can be seen (see Fig. \ref{fig:fsb0} for the 
Fermi surfaces at zero field). The nematic distortion is most obvious in the Fermi surfaces near $(\pm\pi,0)$ and $(0,\pm\pi)$ as 
indicated by the yellow areas in (b). These parts of Fermi surfaces are composed mostly of quasi-1D bands, supporting 
the intimacy of nematic phase to the orbital ordering.
              }
	\label{fig:fschange}
\end{figure*}

The order parameters $\vert\vec{{\cal M}}\vert$ and ${\cal N}$ as 
functions of $\mu_B B$ for this case are shown in Fig. \ref{fig:phase-0}.
There are three rapid increases in the magnetization, consistent 
with the experiment measurements of the real part of the very low
frequency AC magnetic susceptibility at 7.5T, 
7.8T and 8.1T, respectively \cite{grigera2004}.
Experimentally, only the last two exhibit dissipative peaks in 
the imaginary part of the AC susceptibility, which characterize 
the first order metamagnetic transitions.
The first jump measured in  the experiment is considered as a crossover.

The nematic ordering develops in the area bounded by 
the last two magnetization jumps, reproducing the well-known phase 
diagram of the Sr$_3$Ru$_2$O$_7$\cite{grigera2004}.
In particular, if we adopt the results from LDA calculations
\cite{liebsch2000,eremin2002} that $t_1\approx 300$ meV, 
we find that three jumps in the magnetization appear at 
$B\approx 0.0032 t_1/\mu_B$, $0.0053 t_1/\mu_B$, and 
$0.0059 t_1/\mu_B\sim 15.7$ T, 26 T, and 29 T, which is within the
same order to the experimental values of 7.5T, 7.8T, and 8.1T.
This is an improvement compared to the results in previous
theory calculations\cite{kee2005,yamase2007,yamase2007a,
puetter2007,raghu2009,puetter2010}.
in which the nematic ordering develops at much higher field
strength $\mu_B B/t_1\approx 0.02$ 

The sensitivity to the small energy scale like the Zeeman energy is 
because a part of the Fermi surfaces, mostly composed of quasi-1D bands 
as shown in the yellow areas in Fig. \ref{fig:fschange} (b), 
is getting close to the van Hove singularities at $(\pi,0)$ and $(0,\pi)$.
The evolution of the Fermi surface structures as increasing
the B-field across the nematic phase boundaries is presented in
Fig. \ref{fig:fschange} a (before), b (inside), and c (after) 
at $\mu_B B/t_1=0.0048, 0.00544, 0.006$, respectively.
Before and after the nematic phases, the Fermi surfaces have the
4-fold rotational symmetry as exhibited in Fig. \ref{fig:fschange}
(a) and (c).
When the system is in the nematic phase, the Fermi surfaces only 
have 2-fold symmetry as expected.
Particularly, the nematic distortion is most prominent near $(\pm\pi,0)$ 
and $(0,\pm\pi)$ whose Fermi surfaces are dominated by the quasi-1D
bands as shown in Fig. \ref{fig:fschange}(b), supporting the mechanism 
of orbital ordering in quasi-1D bands driven by the van Hove singularities.

It is worthy of mentioning that the onsite spin-orbit coupling
$H_{so}=\lambda \sum_i \vec L_i \cdot \vec S_i$ has important 
effects on the Fermi surface evolutions.
$H_{so}$ hybridizes the opposite spins between quasi-1D bands $d_{yz,xz}$ 
and the 2-D band $d_{xy}$.
As the spin Zeeman energy is present, the spin majority 
(minority) bands of $d_{yz,xz}$ couples to the spin minority (majority) 
band of $d_{xy}$.
Moreover, the orbital Zeeman energy provides more hybridizations between 
quasi-1D $d_{xz,yz}$  bands. 
Combined with the above two effects, the addition of the spin and orbit 
Zeeman energies causes reconstruction of the Fermi surfaces rather
than just chemical potential  shifts.

These results show that the complexity and the sensitivity of the 
Sr$_3$Ru$_2$O$_7$ band structure can be captured very well by our 
tight-binding model with a reasonable quantitative accuracy. 
In the following, the same model will be used to further investigate 
some novel physical properties observed in experiments.

\subsection{Entropy landscape under the magnetic field}
One of the intriguing puzzles observed 
in Sr$_3$Ru$_2$O$_7$ is the critical exponent of the 
divergence in entropy (as well as specific heat) when approaching the 
nematic region, or the quantum critical point. 
Although it is expected that specific heat $C$ diverges as 
$C/T=A[(H-H_c)/H_c]^{-\alpha}+B$ near the quantum critical point, 
the exponent $\alpha$ has been found to be 1\cite{rost2009,rost2010} 
instead of 1/3 as predicted by the Hertz-Millis theory. 
However, it is noticed that ${\cal D}(\epsilon_F)$ is taken to be a 
constant in the Hertz-Millis theory while this 
material has a complicated Fermi surface evolutions under the magnetic field.
Moreover, as mentioned in the introduction, the system is always in the 
Fermi liquid states at low temperatures inferred from the temperature 
dependences of the resistivity.
Motivated by these two facts, it is then worthy of studying first the 
contribution from the band structure to the 
entropy before considering the quantum fluctuations.

\begin{figure}
\includegraphics[width=0.8\linewidth]{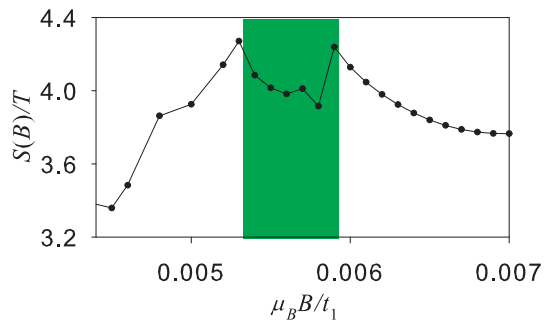}
\caption{\label{fig:entropy} (Color online) 
(a) The entropy landscape represetned by the quantity $S(B)/T$ in unit of $k^2_B/t_1$ within the range of $0.0045\leq \mu_B B/t_1\leq 0.007$.
A sudden increase near the nematic region (green area) is clearly seen.
}
\end{figure}

The entropy per Ru atom at the mean-field level can be evaluated by:
\bea
S(B)= - &\frac{k_B}{N}{\sum_{\kvec}}'\sum_j \big[f(E_j(\vec{k}))\ln f(E_j(\vec{k}))\nn\\
& + (1-f(E_j(\vec{k}))\ln (1-f(E_j(\vec{k})))\big],
\label{eq:entropy}
\eea
where $\beta=1/k_B T$, $f(E_j(\vec{k}))$ the Fermi-Dirac distribution function, and 
$E_j(\vec{k})$ is the $j$th eigenvalue of $H^{MF}(\vec{k})$ given in Eq. \ref{mfall}.
Because only the change of the entropy as a function of magnetic field near the nematic region
is experimentally relevant, we plot in Fig. \ref{fig:entropy} the quantity: $S(B)/T$ in unit of $k^2_B/t_1$
at a low temperature of $1/(\beta t_1)=0.002$ within the range of $0.0045\leq \mu_B B/t_1\leq 0.007$.
Near the nematic region, $S(B)/T$ increases first, being cut off inside the nematic region, and 
then decreases after, which is consistent with the experiment\cite{rost2009}.
Since these results are obtained at the mean-field level, this diverging 
behavior near the nematic region of the entropy landscape results from 
the singularity of ${\cal D}(\epsilon_F)$.
As apparent in Eq. \ref{eq:entropy}, at very low temperature only the 
states near the Fermi surfaces give a sizable contribution to 
the entropy.
Since the nematic region is driven by the sudden increase of 
${\cal D}(\epsilon_F)$, the entropy should also be enhanced near the 
nematic region.
While it is generally expected that the quantum fluctuations near the
critical point could contribute more entropies, our results demonstrate that
entropy contributed from the mean-field level already gives diverging 
behavior due to the singularity of ${\cal D}(\epsilon_F)$ under the 
evolution of the magnetic field.

To summarize, it is found that the singular behavior of 
${\cal D}(\epsilon_F)$ already produces 
diverging behavior in both entropy and specific heat under magnetic field 
at constant temperature, although the critical exponent $\alpha$ can not be extracted from the current theory.
Similar argument has been proposed in previous study\cite{rost2010}, 
in which
the effect of a rigid band shift away from van Hove singularities 
in a perfect 1-d band is discussed.
The quantum fluctuations seem to play a minor role in this case
but become important in the vicinity of the quantum critical point 
and at higher temperature.
It is noted that the rigid band picture does not work neither when the system is doped by the substitution of La$^{3+}$ into the Sr$^{2+}$ site,
which has been argued to result from the strong correlation effect\cite{farrell2008}.
We will leave these discussions in the conclusion section.

\subsection{Total density of states}
Iwaya {\it et al.} has measured the STM tunneling differential conductance
$\frac{dI}{dV}$ in the $B$-field for Sr$_3$Ru$_2$O$_7$, which corresponds 
to measurement of the DOS.
It has been observed that while DOS at higher and lower energy does
not change, the DOS at the Fermi energy (${\cal D}(\epsilon_F)$)
increase significantly under the application of the magnetic field, 
demonstrating the violation of the rigid band picture.

In our model the total DOS can be evaluated using:
\bea
\rho_{tot}(\omega) &=& \frac{1}{\pi N}{\sum_{\vec{k}}}' {\rm Tr}{\rm Im}\big[\hat{G}^{MF}(\vec{k},\omega)\big] \nn \\
\hat{G}^{MF}(\vec{k},\omega)&\equiv& \big(\omega+i\eta- H^{MF}(\vec{k})\big)^{-1}
\label{eq:dos}
\eea
where $H^{MF}(\vec{k})$ is given in Eq. \ref{mfall} with the self-consistent order parameters and $N$ is the total number of sites in the
bilayer square lattices.

\begin{figure}
        \begin{center}
        \subfigure{
        \epsfig{figure=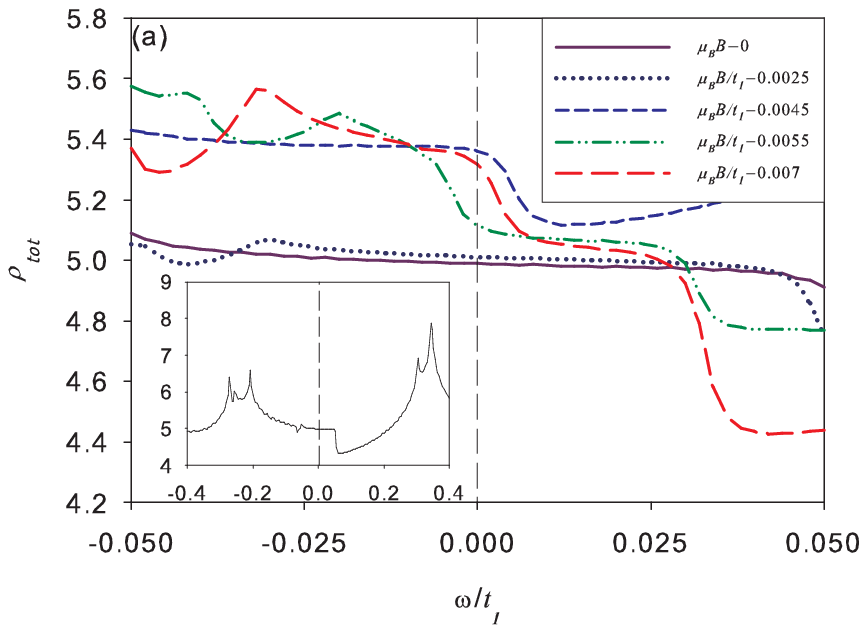,width=0.8\linewidth}}
        \subfigure{
        \epsfig{figure=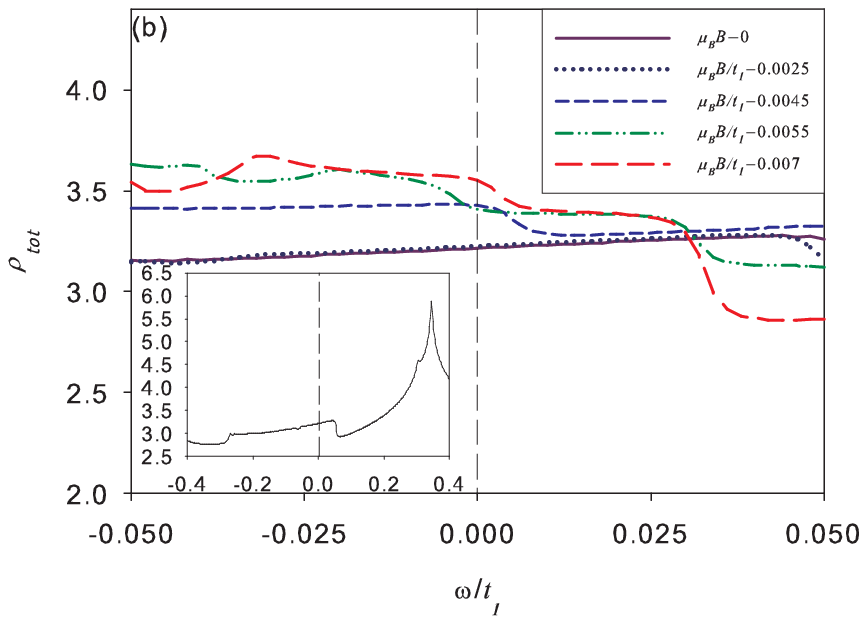,width=0.8\linewidth}}
        \end{center}
        \caption{(Color online) (a) The total DOS $\rho_{tot}(\omega)$ 
as a function of $\mu_B B$. The broadening factor $\eta$ is set to 
be $\eta = 0.002 t_1$.
$\rho_{tot}(\omega)$ does not follow the rigid band picture 
and $\rho_{tot}(\omega=0)$ has a sudden increase near the nematic region. 
Inset:  the total DOS at zero field for a wider range of $\vert\omega\vert/t_1 \leq 0.4$. 
The peaks corresponding to van Hove singularities are near $\omega/t_1\approx -0.2$ and $\omega/t_1\approx 0.35$.
        (b) The DOS of the quasi-1D bands. Inset: the DOS of quasi-1D bands at zero field for a wider range of $\vert\omega\vert/t_1 \leq 0.4$. Only the peak around 
$\omega/t_1\approx 0.35$ remains, meaning that this peak is due to the van Hove singularities in quasi-1D bands.
                }
        \label{fig:dos}
\end{figure}

The profiles of the total DOS at several different magnetic field strength
are plotted in Fig. \ref{fig:dos}(a), and
clearly a rigid band picture does not apply here.
${\cal D}(\epsilon_F)$ ($\rho_{tot}(\omega=0)$ in the plot) increases
significantly as the nematic phase is approached.
This feature can also be directly understood by the picture of Fermi surface reconstruction, since
the changes in the Fermi surface topology inevitably lead to the non-monotonic behavior in $\rho_{tot}(\omega=0)$.
In particular, comparing Figs. \ref{fig:fsb0} and \ref{fig:fschange}, it is straightforward to see that more Fermi surfaces appear near the
$(\pm\pi,0)$ and $(0,\pm\pi)$ as the magnetic field increases.
Since there are van Hove singularities near these four $\vec{k}$ points, $\rho_{tot}(\omega=0)$ is expected to increase.
As the magnetic field is increased further so that the van Hove singularities are all covered below the Fermi surfaces,
$\rho_{tot}(\omega=0)$ starts to drop (not shown here).

At the first glance, the entropy measurement and our results of the 
total DOS seem to contradict with the STM measurement.
While both the entropy measurement and our results develop a maximum
around the nematic region in ${\cal D}(\epsilon_F)$, 
the STM measurement showed instead that ${\cal D}(\epsilon_F)$ keeps 
increasing even after the nematic region is passed.
To resolve this discrepancy, several realistic features need to be
considered before comparing our calculations with the STM results.
Since the STM is a surface probe and the surface of the material is 
usually cleaved to have the oxygen atoms in the outermost layer, 
there is an oxygen atom lying above each uppermost Ru atom.  
Consequently, the tunneling matrix element will be mostly determined by the
wavefunction overlaps between the $p$-orbitals of the oxygen atom and 
the $d$-orbitals of the Ru atom, resulting 
in a much smaller matrix element for $d_{xy}$ orbital compared to 
$d_{xz,yz}$ orbitals\cite{lee2010nematic}.
The minimal model to take this effect into account is to extract the 
DOS only from the quasi-1D orbitals, which is plotted in Fig. \ref{fig:dos}(b).
Although the overall profile in Fig. \ref{fig:dos}(b) is not exactly 
the same as that in Ref. \onlinecite{iwaya2007}, which is attributed to 
more complicated momentum dependence of tunneling matrix 
elements\cite{tersoff1983,zhang2008,lee2009qpi} not considered here, 
it captures the 
increasing DOS with the magnetic field which is more consistent 
with Ref. \onlinecite{iwaya2007}.

The insets in Fig. \ref{fig:dos} plot the total and quasi-1D band DOSs 
at zero field within a wider range of $\vert\omega\vert/t_1 \leq 0.4$.
The peaks corresponding to the van Hove singularities reside at 
$\omega/t_1\approx -0.2$ and $\omega/t_1\approx 0.35$, far away from 
the Fermi energy.
The reason why the small energy scale like Zeeman energy ($\sim 0.003t_1$)
can push the system to get closer to the van Hove singularities at energies 
far away from the Fermi energy is the help of the metamagnetism. 
In the mean-field theory, the magnetization produces an effective chemical potential shift as $s US_z^\alpha$ for electrons at orbital $\alpha$ and spin $s$.
As a result, under the magnetic field the jump in the magnetization
gives $S_z^\alpha\sim 0.05$ within the range of experimental interests.
This leads to the effective chemical potential shift about $\pm 0.18t_1$ 
for $U=3.6t_1$, which is large enough to push the system closer to the
van Hove singularities. 
This renormalization of the chemical potential by the interaction is
also part of the cause for the violation of the rigid band picture.

\subsection{Finite Temperature Phase Diagram}
Another intriguing observation by Rost {\it et. al.}\cite{rost2009} is the phase boundary of the nematic phase at the finite temperature. 
Since the nematic phase is bounded by the two first-order metamagnetic transition, the phase boundary can be determined by the 
magnetic Clausius-Clapeyron relation:
\be
\mu_0\frac{d H_c}{dT_c}=-\frac{\Delta S}{\Delta M},
\label{eq:ccrelation}
\ee
where $\mu_0$ is the permeability constant, $H_c$ and $T_c$ the critical field and temperature, $S$ the entropy, and $M=\vert\vec{{\cal M}}\vert$ the 
magnetization. 
Experimentally it was found that the entropy is always higher inside the nematic phase than the adjacent normal phases but the magnetization increases 
monotonically with the magnetic field.
From Eq. \ref{eq:ccrelation}, this implies a 'muffin'-shaped phase boundary, i.e., at field strengths slightly below $7.8 T$ and above $8.1 T$ the 
nematic phase appears at finite temperature but vanishes at zero temperature, to which we term as 're-entry' behavior throughout this paper.

\begin{figure}
        \begin{center}
        \subfigure{
        \epsfig{figure=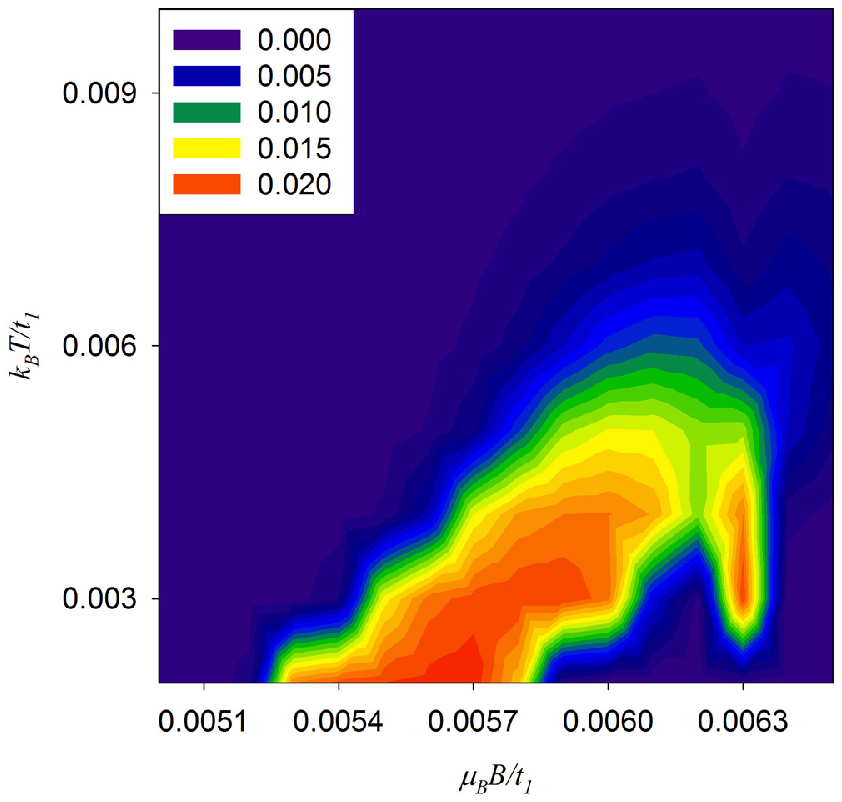,width=0.8\linewidth}}
        \subfigure{
        \epsfig{figure=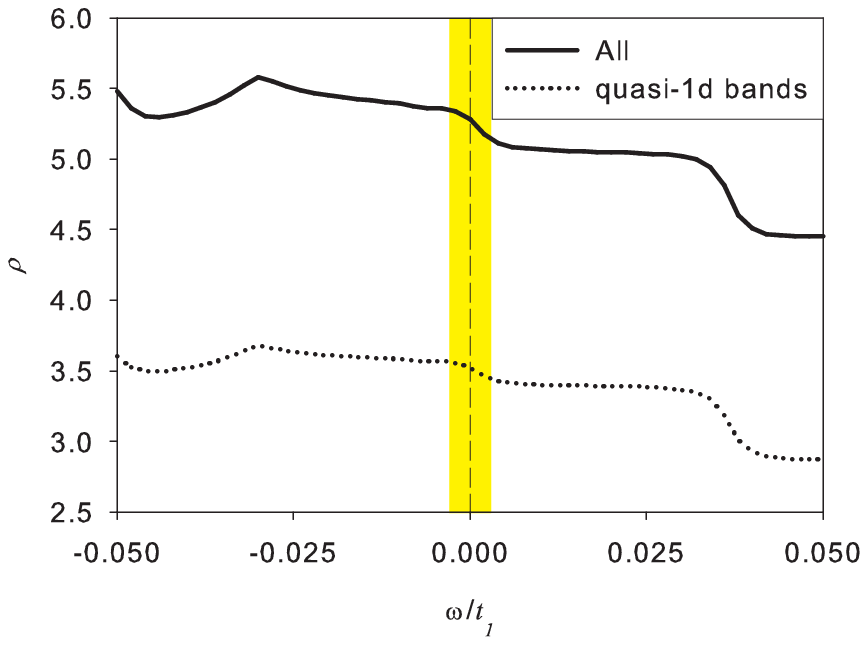,width=0.8\linewidth}}
        \end{center}
        \caption{(Color online) (a) Phase diagram for higher temperatures. Magnitudes of
${\cal N}$ are represented by the color scales. The areas with light colors have large ${\cal N}$, defining the
region for the nematic order. The re-entry of the nematic order at higher temperature is seen at fields between $0.0058<\mu_B B/t_1<0.0063$.
        (b) The DOSs of the all bands (solid line) and quasi-1D bands (dashed line) at $\mu_B B/t_1=0.006$. The yellow areas refers to the energy window bounded by 
$\pm k_B T/t_1$ with temperature $k_B T/t_1=0.003$. It can be seen that this thermal energy window covers a region in which the DOS increases abruptly, 
driving the nematic phase at finite temperature.
                }
        \label{fig:phase-T}
\end{figure}

By inspecting Fig. \ref{fig:entropy} closer, it can be seen that the entropy drops as entering the nematic phase from the lower-field boundary
but raises as entering from the upper-field boundary.
In other words, the present theory has the 're-entry' behavior of the nematic order at the upper-field boundary but not at the lower-field boundary.
This aspect is further confirmed by the temperature dependence of the nematic order parameter ${\cal N}$ as a function of magnetic field shown 
in Fig. \ref{fig:phase-T}(a).

The re-entry behavior can be understood as following. Because the nematic phase transition is first-order, roughly speaking 
one requires $U_{eff}{\cal D}(\epsilon_F)$, where $U_{eff}$ is the effective interaction for nematicity, to exceed a critical value for the nematic ordering to occur.
The mechanism for nematic ordering in Sr$_3$Ru$_2$O$_7$ based on van Hove singularities is all about increasing ${\cal D}(\epsilon_F)$ abruptly 
by driving the system closer to the van Hove singularities with the magnetic field.
At the field strength slightly above the upper critical field, ${\cal D}(\epsilon_F)$ is large but still not enough for the occurrence of the nematic ordering.
Therefore, if the thermal energy is large enough to cover enough DOS within the thermal energy window $\epsilon_F - k_B T<\epsilon_F<\epsilon_F+k_B T$ but 
still low enough so that the thermal fluctuations are small, the re-entry of the nematic phase at higher temperature can be possible. 
As a illustration, Fig. \ref{fig:phase-T}(b) plots the DOS for $\mu_B B/t_1=0.006$ at which the nematic ordering first occurs at $k_B T/t_1=0.003$ 
in our calculation. 
It can be seen that the thermal energy window for $k_B T/t_1=0.003$ (yellow areas) covers a region in which the DOS increases abruptly, 
consistent with the mechanism for the re-entry behavior discussed above.

\section{The case of the tilted magnetic fields ($\theta\neq 0$)}
\label{sect:Btilt}

\subsection{$\theta$ dependence of the nematic ordering}
\begin{figure}
\includegraphics[width=0.8\linewidth]{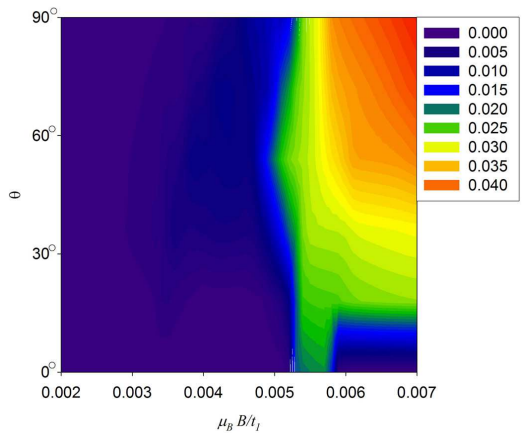}
\caption{\label{fig:phase} (Color online) The nematic order parameter 
${\cal N}$ as functions of $\mu_B B$ and $\theta$.
The magnitudes of ${\cal N}$ are represented by the color scales.
For $\theta\neq 0$, the nematic order parameter is in general non-zero, 
but there is an area (with light colors) in which
the nematic order parameter is sharply enhanced, defining the region of
the nematic phase observed in experiments.}
\end{figure}

Experimentally, it has been concluded from the magnetic susceptibility 
and resistivity measurements\cite{grigera2003,borzi2007,raghu2009} that 
the resistive anisotropy disappears very quickly as the magnetic field 
is tilted away from the $c$-axis, 
suggesting that the nematic ordering vanishes with the increase of 
the field angle $\theta$.
Fig. \ref{fig:phase} presents the nematic order parameter ${\cal N}$ 
as functions of $\mu_B B$ and $\theta$.
It should be noted that since the in-plane component of the orbital
Zeeman energy explicitly break the $C_4$ symmetry down to the $C_2$ symmetry, 
${\cal N}$ is non-zero as long as $\theta\neq 0$.
Nevertheless, the experimentally observable nematic phase can still be 
identified by the jumps in ${\cal N}$.
Our results showed that the nematicity is strongly enhanced 
with the increase of $\theta$, which does not agree with 
the experiments.

\begin{figure}
\includegraphics[width=0.8\linewidth]{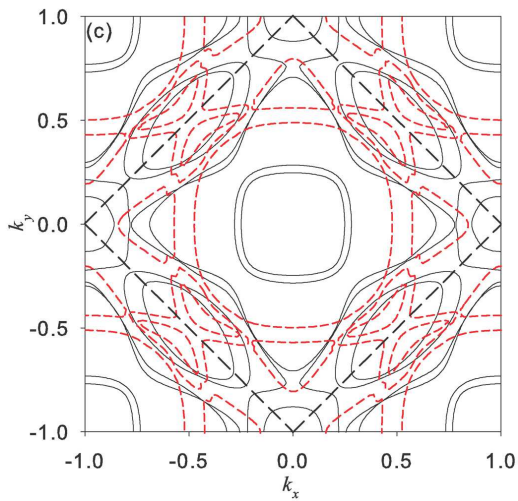}
\caption{\label{fig:fs-90} (Color online) The non-interacting Fermi surfaces for $\vec{B}\parallel \hat{x}$ and $\mu_B B=0.1t_1$.
The anisotropy in Fermi surfaces is already visible, especially for the areas near $(\pm\pi,0)$ and $(0,\pm\pi)$.}
\end{figure}

The large enhancement of the nematic phase for $\theta\neq 0$ 
in our calculation is due to the orbital Zeeman energy.
The anisotropic in-plane component of the orbital Zeeman energy 
term $-\mu_B B \sum_{i,a} L_{x,ia}\sin\theta$ is clearly proportional 
to $B$ and largest at $\theta=90^\circ$ (i.e. $\hat{B}\parallel\hat{x}$). 
To illustrate this point, we plot the Fermi surfaces without any 
interaction for $\vec{B}\parallel \hat{x}$ with strength 
$\mu_B \vert\vec{B}\vert=0.1t_1$ 
in Fig. \ref{fig:fs-90}, and an anisotropy can be seen.
Although such anisotropy in the band structure is not important at low field, 
it can be amplified by the effect of the interactions, driving the system 
more susceptible to the nematic phase as the critical points is approached.
As a result, the portion of the nematic phase in the phase diagram is 
enlarged as $\theta$ increases from $0^\circ$ to $90^\circ$ as seen 
in our calculations.
The disagreement between our theory and the experimentally-determined phase diagram as a function of $\theta$ 
could result from the strong correlation effect, which is beyond the scope of the mean-field theory.

\subsection{Anisotropy of the resistivity measurements}
While it requires more experimental efforts to confirm the cause of disagreement, 
Raghu {\it et. al.} have argued that the resistivity measurement 
might not be a good indicator for the nematic phase in this 
material\cite{raghu2009}. 
The point is that the nematic phase is mostly associated with the states 
near the van Hove singularity whose Fermi velocities are too small to have 
sizable contribution to the transport properties.
Therefore the observed anisotropic resistivity is mostly likely due to 
the scatterings between nematic domains, and the tilt of the magnetic 
field helps lie up the domains.
However, if the domains are fully lied up, the resistivity measurement
becomes insensitive to the nematic phase 
due to the diminishing scatterings between nematic domains,
despite the nematic order could be stronger.

\begin{figure}
\includegraphics[width=0.8\linewidth]{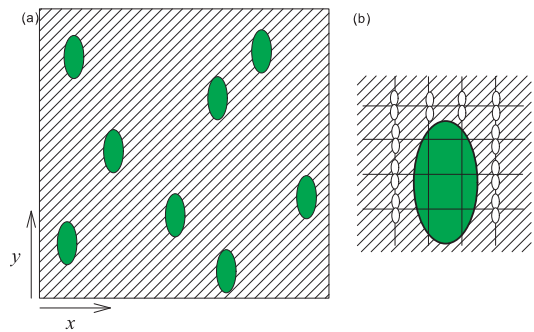}
\caption{\label{fig:domain} (Color online) Illustration of the 
energetically-favored domain structures as the in-plane magnetic 
field is along $\hat{x}$ axis.
When the in-plane magnetic field is weak, orbital ordered phase 
with ${\cal N}>0$ are dominant (shaded areas) and high-energy 
domains with ${\cal N} \leq 0$ (green ovals) could exist. 
(b) The domain walls extend longer in $\hat{y}$ direction because
it costs less energies if less $\hat{y}$- than $\hat{x}$- bonds are broken in 
a ${\cal N}>0$ background. The white oval represents the wave 
function of $d_{yz}$ orbital on each site.
The current flows much more easily along $\hat{y}$-axis than 
$\hat{x}$-axis since the electrons suffer less domain scatterings 
hopping along this direction.
As the in-plane magnetic field is strong, these high-energy domains
vanish, thus there is no longer an easy axis for the current flow.}
\end{figure}


One remaining puzzle of this domain scattering argument is 
why the easy axis for the current flow is perpendicular to 
the in-plane component of the $B$-field \cite{borzi2007}.
We provide a natural argument based on the anisotropic spatial
extension of the domain walls as explained below.
Assuming the $B$-field lying in the $xz$-plane, the in-plane $(xy)$
orbital Zeeman energy reads $H_{in-plane}=-\mu_B B \sin\theta
\sum_{i,a} L_{x,ia}$, which couples the $d_{xy}$ and $d_{xz}$-orbitals and breaks
the degeneracy between the $d_{xz}$ and $d_{yz}$-orbitals.
Since the $d_{xy}$-orbital has lower on-site energy due to the crystal field 
splitting than that of $d_{xz}$, the $d_{xz}$-orbital bands are pushed to 
higher energy than $d_{yz}$-orbital bands by this extra coupling.
As a result, the nematic state with preferred $d_{yz}$-orbitals 
(i.e., ${\cal N}>0$) has lower energy in the homogeneous system. 
At small angles of $\theta$,  domains with preferred $d_{xz}$-orbitals 
(i.e. ${\cal N} \leq 0$) could form as 
depicted in Fig. \ref{fig:domain} (a) as meta-stable states, 
which occupy less volume than the majority domain of ${\cal N}>0$.
Let us consider the shape of the domain walls.
Because of the quasi-1D features of the $d_{xz}$ and $d_{yz}$-orbitals,
the horizontal (vertical) domain wall breaks the bonds of 
the $d_{yz} (d_{xz})$-orbital as depicted 
in Fig. \ref{fig:domain} (b), respectively.
Since the $d_{yz}$-orbital is preferred by $H_{in-plane}$,
the horizontal domain wall costs more energy.
Consequently, the domain structure illustrated in Fig. \ref{fig:domain}
(a) with longer vertical walls than the horizontal walls 
is energetically favored.


Since the electrons suffer less domain scatterings hopping along 
the $\hat{y}$-axis in this domain structure, it becomes the easy axis 
for the current flow.
At large values of $\theta$, higher energy domains are suppressed
and eventually vanish, and thus the resistivity measurement becomes 
insensitive to the nematic phase because of vanishing of the 
domain scatterings.

\subsection{Measurements proposed for the nematicity}
Our results have posted a possibility that the nematic order could
occur in a larger range of the magnetic field for $\vec{B}\parallel 
\hat{x}$ than for $\vec{B}\parallel \hat{z}$.
Detection methods other than resistivity would be desirable.
One feasible way is to measure the 
quasiparticle interference (QPI) in the spectroscopic imaging STM 
which has been examined in detail in our previous work\cite{lee2010nematic}.
It has been predicted by us that if there is a nematic order,
QPI spectra will manifest patterns breaking rotational symmetry.

Another possible experiment is the nuclear quadruple resonance 
(NQR) measurement, which has been widely used to reveal ordered states in 
high-$T_c$ cuprates\cite{hammel1998,teitelbaum2000,singer2002} 
and recently the iron-pnictides\cite{lang2010}.
This technique utilizes the feature that a nucleus with a nuclear 
spin $I\geq 1$ has a non-zero electric quadruple moment. 
Because the electric quadruple moment creates energy splittings in
the nuclear states as a electric field gradient is present, 
a phase transition could be inferred if substantial changes in the 
resonance peak are observed in the NQR measurement.
Besides, since this is a local probe at the atomic level, it is 
highly sensitive to the local electronic change. 
Given that Ru atom has a nuclear spin of $I=5/2$\cite{ishida1997} 
and the orbital ordering in the quasi-1D bands significantly changes 
the charge distribution around the nuclei, 
a systematic NQR measurement as functions of magnetic field and 
field angle will reveal more conclusive information about nematicity.


\section{Conclusions}
\label{sect:cln}

In this paper we have shown that a number of important  properties 
observed in the bilayer Sr$_3$Ru$_2$O$_7$ could be qualitatively consistent with a 
realistic tight-binding model together with on-site interactions 
treated at mean-field level.
The band structure of this material is complicated by multibands, 
bilayer splitting, rotations of RuO octahedra, and the spin-orbit coupling, 
collectively leading to the high sensitivity to the small energy scales. 
This is the main cause of the singular behavior in the evolution of the
Fermi surfaces under magnetic field.
Using our tight-binding model and the standard mean-field approaches 
on the intra- and inter- orbital Hubbard interactions, 
we find that for the case of magnetic field parallel to the $c$-axis, 
the nematic order, which is interpreted as the orbital ordering in 
quais-1d $d_{yz}$ and $d_{xz}$ bands, appears at the magnetic field 
at the same order of the experimental values $\sim$ 8T.
Moreover, we find that the total density of states at the Fermi energy ${\cal D}(\epsilon_F)$ under the magnetic field does not follow a rigid band picture, 
in agreements with the results of STM measurement\cite{iwaya2007}.

The failure of following a rigid band picture is essentially a consequence of the interplay between spin-orbit coupling and the Zeeman energy,
despite the strong correlation effect could also result in the violation of the rigid band shift upon doping\cite{farrell2008}.
Because the spin-orbit coupling hybridizes the quasi-1D bands and 2-D bands with {\it opposite} spins, the Zeeman energy naturally 
induces the reconstruction of the Fermi surfaces instead of just rigid chemical potential shifts.
This singular behavior in ${\cal D}(\epsilon_F)$ also results in the divergences in the entropy and specific heat landscapes, 
since at very low temperature both quantities are approximately proportional to ${\cal D}(\epsilon_F)$. 
Because the divergence observed by Rost {\it el. al.}\cite{rost2009} start approximately at 6-7 Tesla which is not very close to the quantum critical point 
residing about 8 Tesla, a direct application of the quantum critical scaling seems to be inappropriate.
The explanation of the critical exponent associated with this divergence
could not be complete without taking the band structure singularity into
account in this particular material \cite{lee2009qpi327}.

As the magnetic field is tilted away from the $c$ axis ($\theta\neq 0$), 
we find that nematic region expands instead of shrinking 
as the resistivity measurement has indicated.
From the theoretical viewpoints, the tilt of the magnetic field induces 
an extra in-plane component of the orbital Zeeman energy which
explicitly breaks the $C_4$ symmetry down to the $C_2$ symmetry.
As argued above that this system is very sensitive to small energy scale, 
the effect of this extra Zeeman energy is not important at low field but 
could amplify the effect of interaction to drive the system toward 
nematicity as the quantum critical point is approached.
As a result, the nematic phase is more favored and stable in the presence 
of the in-plane magnetic field and it requires another Fermi surface 
reconstruction 
at even higher magnetic field in order to weaken the nematic phase 
by reduced ${\cal D}(\epsilon_F)$.

To explain this discrepancy between our theory and the resistivity measurement, we adopt the domain scattering argument proposed by Raghu {\it et. al.}\cite{raghu2009}. 
Furthermore, we have given an explanation for another experimental puzzle that the easy axis for the current flow is always perpendicular to the 
in-plane magnetic field.
Measurements like quasiparticle interference in the spectroscopic imaging STM and NQR which could detect the orbital ordering directly have been proposed to be 
more reliable probes for the nematicity in this material than the resistivity measurement.

Finally we would like to comment on limitation of the present theory.
Although we have found  the 're-entry' behavior of the nematic phase, i.e., the appearance of the nematic phase only at the finite temperature 
but not at the zero temperature, near the upper-field boundary, the experiments showed this behavior near both upper- and lower- field boundary.
In our calculations, the re-entry behavior is due to the increase of the density of states within the narrow energy window around the Fermi energy opened by thermal 
energy, but we do not reject other schemes for the re-entry behavior.
One possible scheme is an analogue of ferromagnetism without exchange splitting proposed by Hirsch\cite{hirsch1999}.
He showed that the nearest neighbor interactions could result in a spin-dependent renormalization on the bandwidth (equivalently, the effective mass).
As a result, the filling for different spin bands can be different because of the unequal effective masses, leading to the ferromagnetism even without
the exchange splitting as in the Stoner model.

In Hirsch's original paper, the re-entry of the ferromagnetism at higher 
temperature was found.
Since we only considered the on-site interactions in our model, such an effect is beyond the scope of the current theory.
It is possible that after including the nearest neighbor interaction, the renormalizations of the bandwidths have novel temperature-dependences, leading 
to a phase diagram better consistent with the experiments.
If this is the correct scheme, the re-entry of the nematic states should be accompanied by a change in the kinetic energy due to the effective mass renormalization,
which could be examined by the sum rules for the optical properties
\cite{hirsch1999,okimoto1995,okimoto1997,basov2005,laforge2008}.

Another possible scheme for the re-entry behavior is the quantum critical
fluctuations.
It is well-known that the influences of the critical fluctuations extend 
from the quantum critical point to finite temperature in a $V$-shape 
region in the phase diagram.
Moreover, the critical fluctuations in this material contain not only 
the ferromagnetic but also the nematic ones.
As a result, it is not surprising that the competition between these 
two types of critical fluctuations leads to a intriguing phase diagram 
at the finite temperature, 
and the study toward this direction is currently in progress.

\section{Acknowledgment}
We thank J. E. Hirsch and A. Mackenzie for valuable discussions.
C. W. and W. C. L. are supported by ARO-W911NF0810291 and Sloan 
Research Foundation.


\begin{thebibliography}{51}
\expandafter\ifx\csname natexlab\endcsname\relax\def\natexlab#1{#1}\fi
\expandafter\ifx\csname bibnamefont\endcsname\relax
  \def\bibnamefont#1{#1}\fi
\expandafter\ifx\csname bibfnamefont\endcsname\relax
  \def\bibfnamefont#1{#1}\fi
\expandafter\ifx\csname citenamefont\endcsname\relax
  \def\citenamefont#1{#1}\fi
\expandafter\ifx\csname url\endcsname\relax
  \def\url#1{\texttt{#1}}\fi
\expandafter\ifx\csname urlprefix\endcsname\relax\def\urlprefix{URL }\fi
\providecommand{\bibinfo}[2]{#2}
\providecommand{\eprint}[2][]{\url{#2}}

\bibitem[{\citenamefont{Grigera et~al.}(2001)\citenamefont{Grigera, Perry,
  Schofield, Chiao, Julian, Lonzarich, Ikeda, Maeno, Millis, and
  Mackenzie}}]{grigera2001}
\bibinfo{author}{\bibfnamefont{S.~A.} \bibnamefont{Grigera}},
  \bibinfo{author}{\bibfnamefont{R.~S.} \bibnamefont{Perry}},
  \bibinfo{author}{\bibfnamefont{A.~J.} \bibnamefont{Schofield}},
  \bibinfo{author}{\bibfnamefont{M.}~\bibnamefont{Chiao}},
  \bibinfo{author}{\bibfnamefont{S.~R.} \bibnamefont{Julian}},
  \bibinfo{author}{\bibfnamefont{G.~G.} \bibnamefont{Lonzarich}},
  \bibinfo{author}{\bibfnamefont{S.~I.} \bibnamefont{Ikeda}},
  \bibinfo{author}{\bibfnamefont{Y.}~\bibnamefont{Maeno}},
  \bibinfo{author}{\bibfnamefont{A.~J.} \bibnamefont{Millis}},
  \bibnamefont{and} \bibinfo{author}{\bibfnamefont{A.~P.}
  \bibnamefont{Mackenzie}}, \bibinfo{journal}{Science}
  \textbf{\bibinfo{volume}{294}}, \bibinfo{pages}{329} (\bibinfo{year}{2001}).

\bibitem[{\citenamefont{Perry et~al.}(2001)\citenamefont{Perry, Galvin,
  Grigera, Capogna, Schofield, Mackenzie, Chiao, Julian, Ikeda, Nakatsuji
  et~al.}}]{perry2001}
\bibinfo{author}{\bibfnamefont{R.~S.} \bibnamefont{Perry}},
  \bibinfo{author}{\bibfnamefont{L.~M.} \bibnamefont{Galvin}},
  \bibinfo{author}{\bibfnamefont{S.~A.} \bibnamefont{Grigera}},
  \bibinfo{author}{\bibfnamefont{L.}~\bibnamefont{Capogna}},
  \bibinfo{author}{\bibfnamefont{A.~J.} \bibnamefont{Schofield}},
  \bibinfo{author}{\bibfnamefont{A.~P.} \bibnamefont{Mackenzie}},
  \bibinfo{author}{\bibfnamefont{M.}~\bibnamefont{Chiao}},
  \bibinfo{author}{\bibfnamefont{S.~R.} \bibnamefont{Julian}},
  \bibinfo{author}{\bibfnamefont{S.~I.} \bibnamefont{Ikeda}},
  \bibinfo{author}{\bibfnamefont{S.}~\bibnamefont{Nakatsuji}},
  \bibnamefont{et~al.}, \bibinfo{journal}{Phys. Rev. Lett.}
  \textbf{\bibinfo{volume}{86}}, \bibinfo{pages}{2661} (\bibinfo{year}{2001}).

\bibitem[{\citenamefont{Grigera et~al.}(2003)\citenamefont{Grigera, Borzi,
  Mackenzie, Julian, Perry, and Maeno}}]{grigera2003}
\bibinfo{author}{\bibfnamefont{S.~A.} \bibnamefont{Grigera}},
  \bibinfo{author}{\bibfnamefont{R.~A.} \bibnamefont{Borzi}},
  \bibinfo{author}{\bibfnamefont{A.~P.} \bibnamefont{Mackenzie}},
  \bibinfo{author}{\bibfnamefont{S.~R.} \bibnamefont{Julian}},
  \bibinfo{author}{\bibfnamefont{R.~S.} \bibnamefont{Perry}}, \bibnamefont{and}
  \bibinfo{author}{\bibfnamefont{Y.}~\bibnamefont{Maeno}},
  \bibinfo{journal}{Phys. Rev. B} \textbf{\bibinfo{volume}{67}},
  \bibinfo{pages}{214427} (\bibinfo{year}{2003}).

\bibitem[{\citenamefont{Grigera et~al.}(2004)\citenamefont{Grigera, Gegenwart,
  Borzi, Weickert, Schofield, Perry, Tayama, Sakakibara, Maeno, Green
  et~al.}}]{grigera2004}
\bibinfo{author}{\bibfnamefont{S.~A.} \bibnamefont{Grigera}},
  \bibinfo{author}{\bibfnamefont{P.}~\bibnamefont{Gegenwart}},
  \bibinfo{author}{\bibfnamefont{R.~A.} \bibnamefont{Borzi}},
  \bibinfo{author}{\bibfnamefont{F.}~\bibnamefont{Weickert}},
  \bibinfo{author}{\bibfnamefont{A.~J.} \bibnamefont{Schofield}},
  \bibinfo{author}{\bibfnamefont{R.~S.} \bibnamefont{Perry}},
  \bibinfo{author}{\bibfnamefont{T.}~\bibnamefont{Tayama}},
  \bibinfo{author}{\bibfnamefont{T.}~\bibnamefont{Sakakibara}},
  \bibinfo{author}{\bibfnamefont{Y.}~\bibnamefont{Maeno}},
  \bibinfo{author}{\bibfnamefont{A.~G.} \bibnamefont{Green}},
  \bibnamefont{et~al.}, \bibinfo{journal}{Science}
  \textbf{\bibinfo{volume}{306}}, \bibinfo{pages}{1154} (\bibinfo{year}{2004}).

\bibitem[{\citenamefont{Borzi et~al.}(2007)\citenamefont{Borzi, Grigera,
  Farrell, Perry, Lister, Lee, Tennant, Maeno, and Mackenzie}}]{borzi2007}
\bibinfo{author}{\bibfnamefont{R.~A.} \bibnamefont{Borzi}},
  \bibinfo{author}{\bibfnamefont{S.~A.} \bibnamefont{Grigera}},
  \bibinfo{author}{\bibfnamefont{J.}~\bibnamefont{Farrell}},
  \bibinfo{author}{\bibfnamefont{R.~S.} \bibnamefont{Perry}},
  \bibinfo{author}{\bibfnamefont{S.~J.~S.} \bibnamefont{Lister}},
  \bibinfo{author}{\bibfnamefont{S.~L.} \bibnamefont{Lee}},
  \bibinfo{author}{\bibfnamefont{D.~A.} \bibnamefont{Tennant}},
  \bibinfo{author}{\bibfnamefont{Y.}~\bibnamefont{Maeno}}, \bibnamefont{and}
  \bibinfo{author}{\bibfnamefont{A.~P.} \bibnamefont{Mackenzie}},
  \bibinfo{journal}{Science} \textbf{\bibinfo{volume}{315}},
  \bibinfo{pages}{214} (\bibinfo{year}{2007}).

\bibitem[{\citenamefont{Millis et~al.}(2002)\citenamefont{Millis, Schofield,
  Lonzarich, and Grigera}}]{millis2002}
\bibinfo{author}{\bibfnamefont{A.~J.} \bibnamefont{Millis}},
  \bibinfo{author}{\bibfnamefont{A.~J.} \bibnamefont{Schofield}},
  \bibinfo{author}{\bibfnamefont{G.~G.} \bibnamefont{Lonzarich}},
  \bibnamefont{and} \bibinfo{author}{\bibfnamefont{S.~A.}
  \bibnamefont{Grigera}}, \bibinfo{journal}{Phys. Rev. Lett.}
  \textbf{\bibinfo{volume}{88}}, \bibinfo{pages}{217204}
  (\bibinfo{year}{2002}).

\bibitem[{\citenamefont{Green et~al.}(2005)\citenamefont{Green, Grigera, Borzi,
  Mackenzie, Perry, and Simons}}]{green2004}
\bibinfo{author}{\bibfnamefont{A.~G.} \bibnamefont{Green}},
  \bibinfo{author}{\bibfnamefont{S.~A.} \bibnamefont{Grigera}},
  \bibinfo{author}{\bibfnamefont{R.~A.} \bibnamefont{Borzi}},
  \bibinfo{author}{\bibfnamefont{A.~P.} \bibnamefont{Mackenzie}},
  \bibinfo{author}{\bibfnamefont{R.~S.} \bibnamefont{Perry}}, \bibnamefont{and}
  \bibinfo{author}{\bibfnamefont{B.~D.} \bibnamefont{Simons}},
  \bibinfo{journal}{Phys. Rev. Lett.} \textbf{\bibinfo{volume}{95}},
  \bibinfo{pages}{086402} (\bibinfo{year}{2005}).

\bibitem[{\citenamefont{Kee and Kim}(2005)}]{kee2005}
\bibinfo{author}{\bibfnamefont{H.-Y.} \bibnamefont{Kee}} \bibnamefont{and}
  \bibinfo{author}{\bibfnamefont{Y.~B.} \bibnamefont{Kim}},
  \bibinfo{journal}{Phys. Rev. B} \textbf{\bibinfo{volume}{71}},
  \bibinfo{pages}{184402} (\bibinfo{year}{2005}).

\bibitem[{\citenamefont{Yamase and Katanin}(2007)}]{yamase2007}
\bibinfo{author}{\bibfnamefont{H.}~\bibnamefont{Yamase}} \bibnamefont{and}
  \bibinfo{author}{\bibfnamefont{A.}~\bibnamefont{Katanin}},
  \bibinfo{journal}{J. Phys. Soc. Japan} \textbf{\bibinfo{volume}{76}},
  \bibinfo{pages}{073706} (\bibinfo{year}{2007}).

\bibitem[{\citenamefont{Yamase}(2007)}]{yamase2007a}
\bibinfo{author}{\bibfnamefont{H.}~\bibnamefont{Yamase}},
  \bibinfo{journal}{Phys. Rev. B} \textbf{\bibinfo{volume}{76}},
  \bibinfo{pages}{155117} (\bibinfo{year}{2007}).

\bibitem[{\citenamefont{Puetter et~al.}(2007)\citenamefont{Puetter, Doh, and
  Kee}}]{puetter2007}
\bibinfo{author}{\bibfnamefont{C.}~\bibnamefont{Puetter}},
  \bibinfo{author}{\bibfnamefont{H.}~\bibnamefont{Doh}}, \bibnamefont{and}
  \bibinfo{author}{\bibfnamefont{H.-Y.} \bibnamefont{Kee}},
  \bibinfo{journal}{Phys. Rev. B} \textbf{\bibinfo{volume}{76}},
  \bibinfo{pages}{235112} (\bibinfo{year}{2007}).

\bibitem[{\citenamefont{Berridge et~al.}(2009)\citenamefont{Berridge, Green,
  Grigera, and Simons}}]{berridge2009}
\bibinfo{author}{\bibfnamefont{A.~M.} \bibnamefont{Berridge}},
  \bibinfo{author}{\bibfnamefont{A.~G.} \bibnamefont{Green}},
  \bibinfo{author}{\bibfnamefont{S.~A.} \bibnamefont{Grigera}},
  \bibnamefont{and} \bibinfo{author}{\bibfnamefont{B.~D.}
  \bibnamefont{Simons}}, \bibinfo{journal}{Phys. Rev. Lett.}
  \textbf{\bibinfo{volume}{102}}, \bibinfo{pages}{136404}
  (\bibinfo{year}{2009}).

\bibitem[{\citenamefont{Berridge et~al.}(2010)\citenamefont{Berridge, Grigera,
  Simons, and Green}}]{berridge2010}
\bibinfo{author}{\bibfnamefont{A.~M.} \bibnamefont{Berridge}},
  \bibinfo{author}{\bibfnamefont{S.~A.} \bibnamefont{Grigera}},
  \bibinfo{author}{\bibfnamefont{B.~D.} \bibnamefont{Simons}},
  \bibnamefont{and} \bibinfo{author}{\bibfnamefont{A.~G.} \bibnamefont{Green}},
  \bibinfo{journal}{Phys. Rev. B} \textbf{\bibinfo{volume}{81}},
  \bibinfo{pages}{054429} (\bibinfo{year}{2010}).

\bibitem[{\citenamefont{Lee and Wu}(2009{\natexlab{a}})}]{lee2009nematic}
\bibinfo{author}{\bibfnamefont{W.-C.} \bibnamefont{Lee}} \bibnamefont{and}
  \bibinfo{author}{\bibfnamefont{C.}~\bibnamefont{Wu}}, \bibinfo{journal}{Phys.
  Rev. B} \textbf{\bibinfo{volume}{80}}, \bibinfo{pages}{104438}
  (\bibinfo{year}{2009}{\natexlab{a}}).

\bibitem[{\citenamefont{Raghu et~al.}(2009)\citenamefont{Raghu, Paramekanti,
  Kim, Borzi, Grigera, Mackenzie, and Kivelson}}]{raghu2009}
\bibinfo{author}{\bibfnamefont{S.}~\bibnamefont{Raghu}},
  \bibinfo{author}{\bibfnamefont{A.}~\bibnamefont{Paramekanti}},
  \bibinfo{author}{\bibfnamefont{E.~A.} \bibnamefont{Kim}},
  \bibinfo{author}{\bibfnamefont{R.~A.} \bibnamefont{Borzi}},
  \bibinfo{author}{\bibfnamefont{S.~A.} \bibnamefont{Grigera}},
  \bibinfo{author}{\bibfnamefont{A.~P.} \bibnamefont{Mackenzie}},
  \bibnamefont{and} \bibinfo{author}{\bibfnamefont{S.~A.}
  \bibnamefont{Kivelson}}, \bibinfo{journal}{Phys. Rev. B}
  \textbf{\bibinfo{volume}{79}}, \bibinfo{pages}{214402}
  (\bibinfo{year}{2009}).

\bibitem[{\citenamefont{Puetter et~al.}(2010)\citenamefont{Puetter, Rau, and
  Kee}}]{puetter2010}
\bibinfo{author}{\bibfnamefont{C.~M.} \bibnamefont{Puetter}},
  \bibinfo{author}{\bibfnamefont{J.~G.} \bibnamefont{Rau}}, \bibnamefont{and}
  \bibinfo{author}{\bibfnamefont{H.-Y.} \bibnamefont{Kee}},
  \bibinfo{journal}{Phys. Rev. B} \textbf{\bibinfo{volume}{81}},
  \bibinfo{pages}{081105} (\bibinfo{year}{2010}).

\bibitem[{\citenamefont{Fischer and Sigrist}(2010)}]{fischer2010}
\bibinfo{author}{\bibfnamefont{M.~H.} \bibnamefont{Fischer}} \bibnamefont{and}
  \bibinfo{author}{\bibfnamefont{M.}~\bibnamefont{Sigrist}},
  \bibinfo{journal}{Phys. Rev. B} \textbf{\bibinfo{volume}{81}},
  \bibinfo{pages}{064435} (\bibinfo{year}{2010}).

\bibitem[{\citenamefont{Lee et~al.}(2010)\citenamefont{Lee, Arovas, and
  Wu}}]{lee2010nematic}
\bibinfo{author}{\bibfnamefont{W.-C.} \bibnamefont{Lee}},
  \bibinfo{author}{\bibfnamefont{D.~P.} \bibnamefont{Arovas}},
  \bibnamefont{and} \bibinfo{author}{\bibfnamefont{C.}~\bibnamefont{Wu}},
  \bibinfo{journal}{Phys. Rev. B} \textbf{\bibinfo{volume}{81}},
  \bibinfo{pages}{184403} (\bibinfo{year}{2010}).

\bibitem[{\citenamefont{Wu et~al.}(2007)\citenamefont{Wu, Bergman, Balents, and
  Sarma}}]{wu2007}
\bibinfo{author}{\bibfnamefont{C.}~\bibnamefont{Wu}},
  \bibinfo{author}{\bibfnamefont{D.}~\bibnamefont{Bergman}},
  \bibinfo{author}{\bibfnamefont{L.}~\bibnamefont{Balents}}, \bibnamefont{and}
  \bibinfo{author}{\bibfnamefont{S.~D.} \bibnamefont{Sarma}},
  \bibinfo{journal}{Phys. Rev. Lett.} \textbf{\bibinfo{volume}{99}},
  \bibinfo{pages}{70401} (\bibinfo{year}{2007}).

\bibitem[{\citenamefont{Tamai et~al.}(2008)\citenamefont{Tamai, Allan, Mercure,
  Meevasana, Dunkel, Lu, Perry, Mackenzie, Singh, Shen et~al.}}]{tamai2008}
\bibinfo{author}{\bibfnamefont{A.}~\bibnamefont{Tamai}},
  \bibinfo{author}{\bibfnamefont{M.~P.} \bibnamefont{Allan}},
  \bibinfo{author}{\bibfnamefont{J.~F.} \bibnamefont{Mercure}},
  \bibinfo{author}{\bibfnamefont{W.}~\bibnamefont{Meevasana}},
  \bibinfo{author}{\bibfnamefont{R.}~\bibnamefont{Dunkel}},
  \bibinfo{author}{\bibfnamefont{D.~H.} \bibnamefont{Lu}},
  \bibinfo{author}{\bibfnamefont{R.~S.} \bibnamefont{Perry}},
  \bibinfo{author}{\bibfnamefont{A.~P.} \bibnamefont{Mackenzie}},
  \bibinfo{author}{\bibfnamefont{D.~J.} \bibnamefont{Singh}},
  \bibinfo{author}{\bibfnamefont{Z.-X.} \bibnamefont{Shen}},
  \bibnamefont{et~al.}, \bibinfo{journal}{Phys. Rev. Lett.}
  \textbf{\bibinfo{volume}{101}}, \bibinfo{pages}{026407}
  (\bibinfo{year}{2008}).

\bibitem[{\citenamefont{Lee et~al.}(2009)\citenamefont{Lee, Allan, Wang,
  Farrell, Grigera, Baumberger, Davis, and Mackenzie}}]{lee2009qpi327}
\bibinfo{author}{\bibfnamefont{J.}~\bibnamefont{Lee}},
  \bibinfo{author}{\bibfnamefont{M.~P.} \bibnamefont{Allan}},
  \bibinfo{author}{\bibfnamefont{M.~A.} \bibnamefont{Wang}},
  \bibinfo{author}{\bibfnamefont{J.}~\bibnamefont{Farrell}},
  \bibinfo{author}{\bibfnamefont{S.~A.} \bibnamefont{Grigera}},
  \bibinfo{author}{\bibfnamefont{F.}~\bibnamefont{Baumberger}},
  \bibinfo{author}{\bibfnamefont{J.~C.} \bibnamefont{Davis}}, \bibnamefont{and}
  \bibinfo{author}{\bibfnamefont{A.~P.} \bibnamefont{Mackenzie}},
  \bibinfo{journal}{Nat. Phys.} \textbf{\bibinfo{volume}{5}},
  \bibinfo{pages}{800} (\bibinfo{year}{2009}).

\bibitem[{\citenamefont{Salamon and Jaime}(2001)}]{salamon2001}
\bibinfo{author}{\bibfnamefont{M.~B.} \bibnamefont{Salamon}} \bibnamefont{and}
  \bibinfo{author}{\bibfnamefont{M.}~\bibnamefont{Jaime}},
  \bibinfo{journal}{Rev. Mod. Phys.} \textbf{\bibinfo{volume}{73}},
  \bibinfo{pages}{583} (\bibinfo{year}{2001}).

\bibitem[{\citenamefont{Khaliullin}(2005)}]{khaliullin2005}
\bibinfo{author}{\bibfnamefont{G.}~\bibnamefont{Khaliullin}},
  \bibinfo{journal}{Prog. Theor. Phys. Suppl.} \textbf{\bibinfo{volume}{160}},
  \bibinfo{pages}{155} (\bibinfo{year}{2005}).

\bibitem[{\citenamefont{Zhao et~al.}(2009)\citenamefont{Zhao, Adroja, Yao,
  Bewley, Li, Wang, Wu, Chen, Hu,  et~al.}}]{zhao2009}
\bibinfo{author}{\bibfnamefont{J.}~\bibnamefont{Zhao}},
  \bibinfo{author}{\bibfnamefont{D.~T.} \bibnamefont{Adroja}},
  \bibinfo{author}{\bibfnamefont{D.-X.} \bibnamefont{Yao}},
  \bibinfo{author}{\bibfnamefont{R.}~\bibnamefont{Bewley}},
  \bibinfo{author}{\bibfnamefont{S.}~\bibnamefont{Li}},
  \bibinfo{author}{\bibfnamefont{X.~F.} \bibnamefont{Wang}},
  \bibinfo{author}{\bibfnamefont{G.}~\bibnamefont{Wu}},
  \bibinfo{author}{\bibfnamefont{X.~H.} \bibnamefont{Chen}},
  \bibinfo{author}{\bibfnamefont{J.}~\bibnamefont{Hu}}, , \bibnamefont{et~al.},
  \bibinfo{journal}{Nat. Phys.} \textbf{\bibinfo{volume}{5}},
  \bibinfo{pages}{555} (\bibinfo{year}{2009}).

\bibitem[{\citenamefont{Chuang et~al.}(2010)\citenamefont{Chuang, Allan, Lee,
  Xie, Ni, Bud'ko, Boebinger, Canfield, and Davis}}]{chuang2010}
\bibinfo{author}{\bibfnamefont{T.-M.} \bibnamefont{Chuang}},
  \bibinfo{author}{\bibfnamefont{M.~P.} \bibnamefont{Allan}},
  \bibinfo{author}{\bibfnamefont{J.}~\bibnamefont{Lee}},
  \bibinfo{author}{\bibfnamefont{Y.}~\bibnamefont{Xie}},
  \bibinfo{author}{\bibfnamefont{N.}~\bibnamefont{Ni}},
  \bibinfo{author}{\bibfnamefont{S.~L.} \bibnamefont{Bud'ko}},
  \bibinfo{author}{\bibfnamefont{G.~S.} \bibnamefont{Boebinger}},
  \bibinfo{author}{\bibfnamefont{P.~C.} \bibnamefont{Canfield}},
  \bibnamefont{and} \bibinfo{author}{\bibfnamefont{J.~C.} \bibnamefont{Davis}},
  \bibinfo{journal}{Science} \textbf{\bibinfo{volume}{327}},
  \bibinfo{pages}{181} (\bibinfo{year}{2010}).

\bibitem[{\citenamefont{Lee and Wu}(2009{\natexlab{b}})}]{lee2009qpi}
\bibinfo{author}{\bibfnamefont{W.-C.} \bibnamefont{Lee}} \bibnamefont{and}
  \bibinfo{author}{\bibfnamefont{C.}~\bibnamefont{Wu}}, \bibinfo{journal}{Phys.
  Rev. Lett.} \textbf{\bibinfo{volume}{103}}, \bibinfo{pages}{176101}
  (\bibinfo{year}{2009}{\natexlab{b}}).

\bibitem[{\citenamefont{Lv et~al.}(2010)\citenamefont{Lv, Kr\"uger, and
  Phillips}}]{lv2010}
\bibinfo{author}{\bibfnamefont{W.}~\bibnamefont{Lv}},
  \bibinfo{author}{\bibfnamefont{F.}~\bibnamefont{Kr\"uger}}, \bibnamefont{and}
  \bibinfo{author}{\bibfnamefont{P.}~\bibnamefont{Phillips}},
  \bibinfo{journal}{Phys. Rev. B} \textbf{\bibinfo{volume}{82}},
  \bibinfo{pages}{045125} (\bibinfo{year}{2010}).

\bibitem[{\citenamefont{{Chen} et~al.}(2010)\citenamefont{{Chen}, {Maciejko},
  {Sorini}, {Moritz}, {Singh}, and {Devereaux}}}]{chen2010}
\bibinfo{author}{\bibfnamefont{C.}~\bibnamefont{{Chen}}},
  \bibinfo{author}{\bibfnamefont{J.}~\bibnamefont{{Maciejko}}},
  \bibinfo{author}{\bibfnamefont{A.~P.} \bibnamefont{{Sorini}}},
  \bibinfo{author}{\bibfnamefont{B.}~\bibnamefont{{Moritz}}},
  \bibinfo{author}{\bibfnamefont{R.~R.~P.} \bibnamefont{{Singh}}},
  \bibnamefont{and} \bibinfo{author}{\bibfnamefont{T.~P.}
  \bibnamefont{{Devereaux}}}, \bibinfo{howpublished}{arXiv.org:1004.4611}
  (\bibinfo{year}{2010}).

\bibitem[{\citenamefont{Coleman and Schofield}(2005)}]{coleman2005}
\bibinfo{author}{\bibfnamefont{P.}~\bibnamefont{Coleman}} \bibnamefont{and}
  \bibinfo{author}{\bibfnamefont{A.~J.} \bibnamefont{Schofield}},
  \bibinfo{journal}{Nature} \textbf{\bibinfo{volume}{433}},
  \bibinfo{pages}{226} (\bibinfo{year}{2005}).

\bibitem[{\citenamefont{L\"ohneysen et~al.}(2007)\citenamefont{L\"ohneysen,
  Rosch, Vojta, and W\"olfle}}]{lohneysen2007}
\bibinfo{author}{\bibfnamefont{H.~v.} \bibnamefont{L\"ohneysen}},
  \bibinfo{author}{\bibfnamefont{A.}~\bibnamefont{Rosch}},
  \bibinfo{author}{\bibfnamefont{M.}~\bibnamefont{Vojta}}, \bibnamefont{and}
  \bibinfo{author}{\bibfnamefont{P.}~\bibnamefont{W\"olfle}},
  \bibinfo{journal}{Rev. Mod. Phys.} \textbf{\bibinfo{volume}{79}},
  \bibinfo{pages}{1015} (\bibinfo{year}{2007}).

\bibitem[{\citenamefont{Hertz}(1976)}]{hertz1976}
\bibinfo{author}{\bibfnamefont{J.~A.} \bibnamefont{Hertz}},
  \bibinfo{journal}{Phys. Rev. B} \textbf{\bibinfo{volume}{14}},
  \bibinfo{pages}{1165} (\bibinfo{year}{1976}).

\bibitem[{\citenamefont{Millis}(1993)}]{millis1993}
\bibinfo{author}{\bibfnamefont{A.~J.} \bibnamefont{Millis}},
  \bibinfo{journal}{Phys. Rev. B} \textbf{\bibinfo{volume}{48}},
  \bibinfo{pages}{7183} (\bibinfo{year}{1993}).

\bibitem[{\citenamefont{Rost et~al.}(2009)\citenamefont{Rost, Perry, Mercure,
  Mackenzie, and Grigera}}]{rost2009}
\bibinfo{author}{\bibfnamefont{A.~W.} \bibnamefont{Rost}},
  \bibinfo{author}{\bibfnamefont{R.~S.} \bibnamefont{Perry}},
  \bibinfo{author}{\bibfnamefont{J.-F.} \bibnamefont{Mercure}},
  \bibinfo{author}{\bibfnamefont{A.~P.} \bibnamefont{Mackenzie}},
  \bibnamefont{and} \bibinfo{author}{\bibfnamefont{S.~A.}
  \bibnamefont{Grigera}}, \bibinfo{journal}{Science}
  \textbf{\bibinfo{volume}{325}}, \bibinfo{pages}{1360} (\bibinfo{year}{2009}).

\bibitem[{\citenamefont{Rost et~al.}(2010)\citenamefont{Rost, Berridge, Perry,
  Mercure, Grigera, and Mackenzie}}]{rost2010}
\bibinfo{author}{\bibfnamefont{A.~W.} \bibnamefont{Rost}},
  \bibinfo{author}{\bibfnamefont{A.~M.} \bibnamefont{Berridge}},
  \bibinfo{author}{\bibfnamefont{R.~S.} \bibnamefont{Perry}},
  \bibinfo{author}{\bibfnamefont{J.-F.} \bibnamefont{Mercure}},
  \bibinfo{author}{\bibfnamefont{S.~A.} \bibnamefont{Grigera}},
  \bibnamefont{and} \bibinfo{author}{\bibfnamefont{A.~P.}
  \bibnamefont{Mackenzie}}, \bibinfo{journal}{Phys. Status Solidi B}
  \textbf{\bibinfo{volume}{247}}, \bibinfo{pages}{513} (\bibinfo{year}{2010}).

\bibitem[{\citenamefont{Iwaya et~al.}(2007)\citenamefont{Iwaya, Satow,
  Hanaguri, Shannon, Yoshida, Ikeda, He, Kaneko, Tokura, Yamada
  et~al.}}]{iwaya2007}
\bibinfo{author}{\bibfnamefont{K.}~\bibnamefont{Iwaya}},
  \bibinfo{author}{\bibfnamefont{S.}~\bibnamefont{Satow}},
  \bibinfo{author}{\bibfnamefont{T.}~\bibnamefont{Hanaguri}},
  \bibinfo{author}{\bibfnamefont{N.}~\bibnamefont{Shannon}},
  \bibinfo{author}{\bibfnamefont{Y.}~\bibnamefont{Yoshida}},
  \bibinfo{author}{\bibfnamefont{S.~I.} \bibnamefont{Ikeda}},
  \bibinfo{author}{\bibfnamefont{J.~P.} \bibnamefont{He}},
  \bibinfo{author}{\bibfnamefont{Y.}~\bibnamefont{Kaneko}},
  \bibinfo{author}{\bibfnamefont{Y.}~\bibnamefont{Tokura}},
  \bibinfo{author}{\bibfnamefont{T.}~\bibnamefont{Yamada}},
  \bibnamefont{et~al.}, \bibinfo{journal}{Phys. Rev. Lett.}
  \textbf{\bibinfo{volume}{99}}, \bibinfo{pages}{057208}
  (\bibinfo{year}{2007}).

\bibitem[{\citenamefont{Singh and Mazin}(2001)}]{singh2001}
\bibinfo{author}{\bibfnamefont{D.}~\bibnamefont{Singh}} \bibnamefont{and}
  \bibinfo{author}{\bibfnamefont{I.}~\bibnamefont{Mazin}},
  \bibinfo{journal}{Phys. Rev. B} \textbf{\bibinfo{volume}{63}},
  \bibinfo{pages}{165101} (\bibinfo{year}{2001}).

\bibitem[{\citenamefont{Liebsch and Lichtenstein}(2000)}]{liebsch2000}
\bibinfo{author}{\bibfnamefont{A.}~\bibnamefont{Liebsch}} \bibnamefont{and}
  \bibinfo{author}{\bibfnamefont{A.}~\bibnamefont{Lichtenstein}},
  \bibinfo{journal}{Phys. Rev. Lett.} \textbf{\bibinfo{volume}{84}},
  \bibinfo{pages}{1591} (\bibinfo{year}{2000}).

\bibitem[{\citenamefont{Eremin et~al.}(2002)\citenamefont{Eremin, Manske, and
  Bennemann}}]{eremin2002}
\bibinfo{author}{\bibfnamefont{I.}~\bibnamefont{Eremin}},
  \bibinfo{author}{\bibfnamefont{D.}~\bibnamefont{Manske}}, \bibnamefont{and}
  \bibinfo{author}{\bibfnamefont{K.}~\bibnamefont{Bennemann}},
  \bibinfo{journal}{Phys. Rev. B} \textbf{\bibinfo{volume}{65}},
  \bibinfo{pages}{220502} (\bibinfo{year}{2002}).

\bibitem[{\citenamefont{Farrell et~al.}(2008)\citenamefont{Farrell, Perry,
  Rost, Mercure, Kikugawa, Grigera, and Mackenzie}}]{farrell2008}
\bibinfo{author}{\bibfnamefont{J.}~\bibnamefont{Farrell}},
  \bibinfo{author}{\bibfnamefont{R.~S.} \bibnamefont{Perry}},
  \bibinfo{author}{\bibfnamefont{A.}~\bibnamefont{Rost}},
  \bibinfo{author}{\bibfnamefont{J.~F.} \bibnamefont{Mercure}},
  \bibinfo{author}{\bibfnamefont{N.}~\bibnamefont{Kikugawa}},
  \bibinfo{author}{\bibfnamefont{S.~A.} \bibnamefont{Grigera}},
  \bibnamefont{and} \bibinfo{author}{\bibfnamefont{A.~P.}
  \bibnamefont{Mackenzie}}, \bibinfo{journal}{Phys. Rev. B}
  \textbf{\bibinfo{volume}{78}}, \bibinfo{pages}{180409}
  (\bibinfo{year}{2008}).

\bibitem[{\citenamefont{Tersoff and Hamann}(1983)}]{tersoff1983}
\bibinfo{author}{\bibfnamefont{J.}~\bibnamefont{Tersoff}} \bibnamefont{and}
  \bibinfo{author}{\bibfnamefont{D.}~\bibnamefont{Hamann}},
  \bibinfo{journal}{Phys. Rev. Lett.} \textbf{\bibinfo{volume}{50}},
  \bibinfo{pages}{1998} (\bibinfo{year}{1983}).

\bibitem[{\citenamefont{Zhang et~al.}(2008)\citenamefont{Zhang, Brar, Wang,
  Girit, Yayon, Panlasigui, Zettl, and Crommie}}]{zhang2008}
\bibinfo{author}{\bibfnamefont{Y.}~\bibnamefont{Zhang}},
  \bibinfo{author}{\bibfnamefont{V.~W.} \bibnamefont{Brar}},
  \bibinfo{author}{\bibfnamefont{F.}~\bibnamefont{Wang}},
  \bibinfo{author}{\bibfnamefont{C.}~\bibnamefont{Girit}},
  \bibinfo{author}{\bibfnamefont{Y.}~\bibnamefont{Yayon}},
  \bibinfo{author}{\bibfnamefont{M.}~\bibnamefont{Panlasigui}},
  \bibinfo{author}{\bibfnamefont{A.}~\bibnamefont{Zettl}}, \bibnamefont{and}
  \bibinfo{author}{\bibfnamefont{M.~F.} \bibnamefont{Crommie}},
  \bibinfo{journal}{Nat. Phys.} \textbf{\bibinfo{volume}{4}},
  \bibinfo{pages}{627} (\bibinfo{year}{2008}).

\bibitem[{\citenamefont{Hammel et~al.}(1998)\citenamefont{Hammel, Statt,
  Martin, Chou, Johnston, and Cheong}}]{hammel1998}
\bibinfo{author}{\bibfnamefont{P.~C.} \bibnamefont{Hammel}},
  \bibinfo{author}{\bibfnamefont{B.~W.} \bibnamefont{Statt}},
  \bibinfo{author}{\bibfnamefont{R.~L.} \bibnamefont{Martin}},
  \bibinfo{author}{\bibfnamefont{F.~C.} \bibnamefont{Chou}},
  \bibinfo{author}{\bibfnamefont{D.~C.} \bibnamefont{Johnston}},
  \bibnamefont{and} \bibinfo{author}{\bibfnamefont{S.-W.}
  \bibnamefont{Cheong}}, \bibinfo{journal}{Phys. Rev. B}
  \textbf{\bibinfo{volume}{57}}, \bibinfo{pages}{R712} (\bibinfo{year}{1998}).

\bibitem[{\citenamefont{Teitel'baum et~al.}(2000)\citenamefont{Teitel'baum,
  B\"uchner, and de~Gronckel}}]{teitelbaum2000}
\bibinfo{author}{\bibfnamefont{G.~B.} \bibnamefont{Teitel'baum}},
  \bibinfo{author}{\bibfnamefont{B.}~\bibnamefont{B\"uchner}},
  \bibnamefont{and}
  \bibinfo{author}{\bibfnamefont{H.}~\bibnamefont{de~Gronckel}},
  \bibinfo{journal}{Phys. Rev. Lett.} \textbf{\bibinfo{volume}{84}},
  \bibinfo{pages}{2949} (\bibinfo{year}{2000}).

\bibitem[{\citenamefont{Singer et~al.}(2002)\citenamefont{Singer, Hunt, and
  Imai}}]{singer2002}
\bibinfo{author}{\bibfnamefont{P.~M.} \bibnamefont{Singer}},
  \bibinfo{author}{\bibfnamefont{A.~W.} \bibnamefont{Hunt}}, \bibnamefont{and}
  \bibinfo{author}{\bibfnamefont{T.}~\bibnamefont{Imai}},
  \bibinfo{journal}{Phys. Rev. Lett.} \textbf{\bibinfo{volume}{88}},
  \bibinfo{pages}{047602} (\bibinfo{year}{2002}).

\bibitem[{\citenamefont{Lang et~al.}(2010)\citenamefont{Lang, Grafe, Paar,
  Hammerath, Manthey, Behr, Werner, and B\"uchner}}]{lang2010}
\bibinfo{author}{\bibfnamefont{G.}~\bibnamefont{Lang}},
  \bibinfo{author}{\bibfnamefont{H.-J.} \bibnamefont{Grafe}},
  \bibinfo{author}{\bibfnamefont{D.}~\bibnamefont{Paar}},
  \bibinfo{author}{\bibfnamefont{F.}~\bibnamefont{Hammerath}},
  \bibinfo{author}{\bibfnamefont{K.}~\bibnamefont{Manthey}},
  \bibinfo{author}{\bibfnamefont{G.}~\bibnamefont{Behr}},
  \bibinfo{author}{\bibfnamefont{J.}~\bibnamefont{Werner}}, \bibnamefont{and}
  \bibinfo{author}{\bibfnamefont{B.}~\bibnamefont{B\"uchner}},
  \bibinfo{journal}{Phys. Rev. Lett.} \textbf{\bibinfo{volume}{104}},
  \bibinfo{pages}{097001} (\bibinfo{year}{2010}).

\bibitem[{\citenamefont{Ishida et~al.}(1997)\citenamefont{Ishida, Kitaoka,
  Asayama, Ikeda, Nishizaki, Maeno, Yoshida, and Fujita}}]{ishida1997}
\bibinfo{author}{\bibfnamefont{K.}~\bibnamefont{Ishida}},
  \bibinfo{author}{\bibfnamefont{Y.}~\bibnamefont{Kitaoka}},
  \bibinfo{author}{\bibfnamefont{K.}~\bibnamefont{Asayama}},
  \bibinfo{author}{\bibfnamefont{S.}~\bibnamefont{Ikeda}},
  \bibinfo{author}{\bibfnamefont{S.}~\bibnamefont{Nishizaki}},
  \bibinfo{author}{\bibfnamefont{Y.}~\bibnamefont{Maeno}},
  \bibinfo{author}{\bibfnamefont{K.}~\bibnamefont{Yoshida}}, \bibnamefont{and}
  \bibinfo{author}{\bibfnamefont{T.}~\bibnamefont{Fujita}},
  \bibinfo{journal}{Phys. Rev. B} \textbf{\bibinfo{volume}{56}},
  \bibinfo{pages}{R505} (\bibinfo{year}{1997}).

\bibitem[{\citenamefont{Hirsch}(1999)}]{hirsch1999}
\bibinfo{author}{\bibfnamefont{J.~E.} \bibnamefont{Hirsch}},
  \bibinfo{journal}{Phys. Rev. B} \textbf{\bibinfo{volume}{59}},
  \bibinfo{pages}{6256} (\bibinfo{year}{1999}).

\bibitem[{\citenamefont{Okimoto et~al.}(1995)\citenamefont{Okimoto, Katsufuji,
  Ishikawa, Urushibara, Arima, and Tokura}}]{okimoto1995}
\bibinfo{author}{\bibfnamefont{Y.}~\bibnamefont{Okimoto}},
  \bibinfo{author}{\bibfnamefont{T.}~\bibnamefont{Katsufuji}},
  \bibinfo{author}{\bibfnamefont{T.}~\bibnamefont{Ishikawa}},
  \bibinfo{author}{\bibfnamefont{A.}~\bibnamefont{Urushibara}},
  \bibinfo{author}{\bibfnamefont{T.}~\bibnamefont{Arima}}, \bibnamefont{and}
  \bibinfo{author}{\bibfnamefont{Y.}~\bibnamefont{Tokura}},
  \bibinfo{journal}{Phys. Rev. Lett.} \textbf{\bibinfo{volume}{75}},
  \bibinfo{pages}{109} (\bibinfo{year}{1995}).

\bibitem[{\citenamefont{Okimoto et~al.}(1997)\citenamefont{Okimoto, Katsufuji,
  Ishikawa, Arima, and Tokura}}]{okimoto1997}
\bibinfo{author}{\bibfnamefont{Y.}~\bibnamefont{Okimoto}},
  \bibinfo{author}{\bibfnamefont{T.}~\bibnamefont{Katsufuji}},
  \bibinfo{author}{\bibfnamefont{T.}~\bibnamefont{Ishikawa}},
  \bibinfo{author}{\bibfnamefont{T.}~\bibnamefont{Arima}}, \bibnamefont{and}
  \bibinfo{author}{\bibfnamefont{Y.}~\bibnamefont{Tokura}},
  \bibinfo{journal}{Phys. Rev. B} \textbf{\bibinfo{volume}{55}},
  \bibinfo{pages}{4206} (\bibinfo{year}{1997}).

\bibitem[{\citenamefont{Basov and Timusk}(2005)}]{basov2005}
\bibinfo{author}{\bibfnamefont{D.~N.} \bibnamefont{Basov}} \bibnamefont{and}
  \bibinfo{author}{\bibfnamefont{T.}~\bibnamefont{Timusk}},
  \bibinfo{journal}{Rev. Mod. Phys.} \textbf{\bibinfo{volume}{77}},
  \bibinfo{pages}{721} (\bibinfo{year}{2005}).

\bibitem[{\citenamefont{LaForge et~al.}(2008)\citenamefont{LaForge, Padilla,
  Burch, Li, Schafgans, Segawa, Ando, and Basov}}]{laforge2008}
\bibinfo{author}{\bibfnamefont{A.~D.} \bibnamefont{LaForge}},
  \bibinfo{author}{\bibfnamefont{W.~J.} \bibnamefont{Padilla}},
  \bibinfo{author}{\bibfnamefont{K.~S.} \bibnamefont{Burch}},
  \bibinfo{author}{\bibfnamefont{Z.~Q.} \bibnamefont{Li}},
  \bibinfo{author}{\bibfnamefont{A.~A.} \bibnamefont{Schafgans}},
  \bibinfo{author}{\bibfnamefont{K.}~\bibnamefont{Segawa}},
  \bibinfo{author}{\bibfnamefont{Y.}~\bibnamefont{Ando}}, \bibnamefont{and}
  \bibinfo{author}{\bibfnamefont{D.~N.} \bibnamefont{Basov}},
  \bibinfo{journal}{Phys. Rev. Lett.} \textbf{\bibinfo{volume}{101}},
  \bibinfo{pages}{097008} (\bibinfo{year}{2008}).

\end{thebibliography}

\end{document}